\documentclass[twoside,twocolumn,9pt]{article}
\usepackage{extsizes}
\usepackage[super,sort&compress,comma]{natbib} 
\usepackage[version=3]{mhchem}
\usepackage[left=1.5cm, right=1.5cm, top=1.785cm, bottom=2.0cm]{geometry}
\usepackage{balance}
\usepackage{mathptmx}
\usepackage{sectsty}
\usepackage{graphicx} 
\usepackage{lastpage}
\usepackage[format=plain,justification=justified,singlelinecheck=false,font={stretch=1.125,small,sf},labelfont=bf,labelsep=space]{caption}
\usepackage{float}
\usepackage{fancyhdr}
\usepackage{fnpos}
\usepackage[english]{babel}
\addto{\captionsenglish}{%
  
}
\usepackage{array}
\usepackage{droidsans}
\usepackage{charter}
\usepackage[T1]{fontenc}
\usepackage[usenames,dvipsnames]{xcolor}
\usepackage{setspace}
\usepackage[compact]{titlesec}
\usepackage{hyperref}

\usepackage{epstopdf}

\usepackage[switch]{lineno}
\usepackage{multirow}

\newcommand{\rev}[1]{\textcolor{black}{#1}}

\definecolor{cream}{RGB}{222,217,201}

\begin{document}

\pagestyle{fancy}
\thispagestyle{plain}
\fancypagestyle{plain}{
\renewcommand{\headrulewidth}{0pt}
}

\makeFNbottom
\makeatletter
\renewcommand\LARGE{\@setfontsize\LARGE{15pt}{17}}
\renewcommand\Large{\@setfontsize\Large{12pt}{14}}
\renewcommand\large{\@setfontsize\large{10pt}{12}}
\renewcommand\footnotesize{\@setfontsize\footnotesize{7pt}{10}}
\makeatother

\renewcommand{\thefootnote}{\fnsymbol{footnote}}
\renewcommand\footnoterule{\vspace*{1pt}%
\color{cream}\hrule width 3.5in height 0.4pt \color{black}\vspace*{5pt}} 
\setcounter{secnumdepth}{5}

\makeatletter 
\renewcommand\@biblabel[1]{#1}            
\renewcommand\@makefntext[1]%
{\noindent\makebox[0pt][r]{\@thefnmark\,}#1}
\makeatother 
\renewcommand{\figurename}{\small{Fig.}~}
\sectionfont{\sffamily\Large}
\subsectionfont{\normalsize}
\subsubsectionfont{\bf}
\setstretch{1.125} 
\setlength{\skip\footins}{0.8cm}
\setlength{\footnotesep}{0.25cm}
\setlength{\jot}{10pt}
\titlespacing*{\section}{0pt}{4pt}{4pt}
\titlespacing*{\subsection}{0pt}{15pt}{1pt}

\fancyfoot{}
\fancyfoot[LO,RE]{\vspace{-7.1pt}\includegraphics[height=9pt]{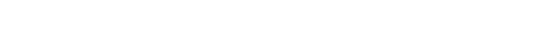}}
\fancyfoot[CO]{\vspace{-7.1pt}\hspace{13.2cm}\includegraphics{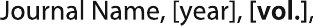}}
\fancyfoot[CE]{\vspace{-7.2pt}\hspace{-14.2cm}\includegraphics{head_foot/RF}}
\fancyfoot[RO]{\footnotesize{\sffamily{1--\pageref{LastPage} ~\textbar  \hspace{2pt}\thepage}}}
\fancyfoot[LE]{\footnotesize{\sffamily{\thepage~\textbar\hspace{3.45cm} 1--\pageref{LastPage}}}}
\fancyhead{}
\renewcommand{\headrulewidth}{0pt} 
\renewcommand{\footrulewidth}{0pt}
\setlength{\arrayrulewidth}{1pt}
\setlength{\columnsep}{6.5mm}
\setlength\bibsep{1pt}

\makeatletter 
\newlength{\figrulesep} 
\setlength{\figrulesep}{0.5\textfloatsep} 

\newcommand{\topfigrule}{\vspace*{-1pt}%
\noindent{\color{cream}\rule[-\figrulesep]{\columnwidth}{1.5pt}} }

\newcommand{\botfigrule}{\vspace*{-2pt}%
\noindent{\color{cream}\rule[\figrulesep]{\columnwidth}{1.5pt}} }

\newcommand{\dblfigrule}{\vspace*{-1pt}%
\noindent{\color{cream}\rule[-\figrulesep]{\textwidth}{1.5pt}} }

\makeatother

\twocolumn[
  \begin{@twocolumnfalse}
{\includegraphics[height=30pt]{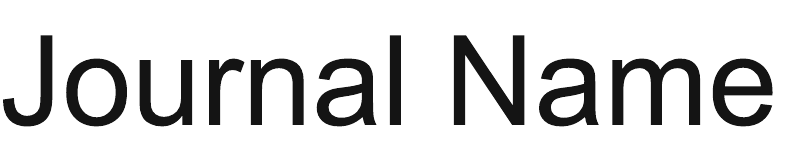}\hfill\raisebox{0pt}[0pt][0pt]{\includegraphics[height=55pt]{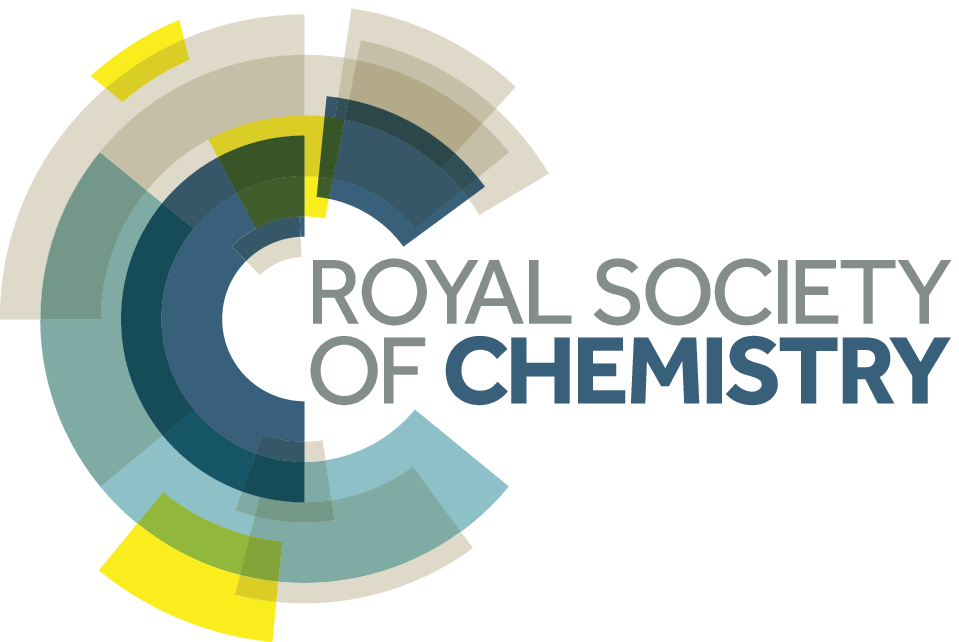}}\\[1ex]
\includegraphics[width=18.5cm]{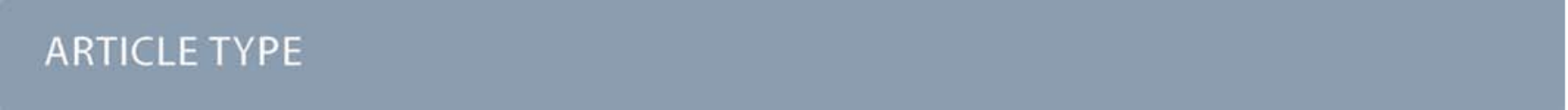}}\par
\vspace{1em}
\sffamily
\begin{tabular}{m{4.5cm} p{13.5cm} }

\includegraphics{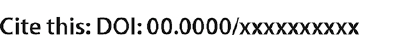} & \noindent\LARGE{\textbf{Tempering of Au nanoclusters: capturing the temperature-dependent competition among structural motifs$^\dag$}} \\
\vspace{0.3cm} & \vspace{0.3cm} \\

 & \noindent\large{Manoj Settem,\textit{$^{a}$} Riccardo Ferrando,$^{\ast}$\textit{$^{b}$} and Alberto Giacomello$^{\ast}$\textit{$^{a}$}} \\

\includegraphics{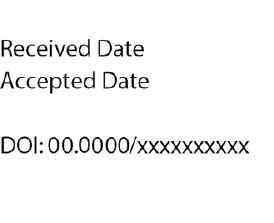} & 
\noindent\normalsize{A computational approach to determine the equilibrium structures of nanoclusters in the whole temperature range from 0 K to melting is developed. Our approach relies on Parallel Tempering Molecular Dynamics (PTMD) simulations complemented by Harmonic Superposition Approximation (HSA) calculations and global optimization searches, thus combining the accuracy of global optimization and HSA in describing the low-energy part of configuration space, together with the PTDM thorough sampling of high-energy configurations.
This combined methodology is shown to be instrumental towards revealing the temperature-dependent structural motifs in Au nanoclusters of sizes 90, 147, and 201 atoms. The reported phenomenology is particularly rich, displaying a size- and temperature-dependent competition between the global energy minimum and other structural motifs. In the case of Au$_{90}$ and Au$_{147}$, the global minimum is also the dominant structure at finite temperatures. In contrast, the Au$_{201}$ cluster undergoes a solid-solid transformation at low temperature (< 200 K). Results indicate that PTMD and HSA very well agree at intermediate temperatures, between 300 and 400 K. For higher temperatures, PTMD gives an accurate description of equilibrium, while HSA fails in describing the melting range. On the other hand, HSA is more efficient in catching low-temperature structural transitions. Finally, we describe the elusive structures close to the melting region which can present complex and defective geometries, that are otherwise difficult to characterize through experimental imaging.} \\

\end{tabular}

 \end{@twocolumnfalse} \vspace{0.6cm}

  ]

\renewcommand*\rmdefault{bch}\normalfont\upshape
\rmfamily
\section*{}
\vspace{-1cm}


\footnotetext{\textit{$^{a}$~Dipartimento di Ingegneria Meccanica e Aerospaziale, Sapienza Università di Roma, via Eudossiana 18, 00184 Roma, Italy. E-mail: alberto.giacomello@uniroma1.it}}
\footnotetext{\textit{$^{b}$~Dipartimento di Fisica dell'Università di Genova and CNR-IMEM, via Dodecaneso 33, 16146 Genova, Italy. E-mail: ferrando@fisica.unige.it}}

\footnotetext{\dag~Electronic Supplementary Information (ESI) available: [ESI contains the results of PTMD simulations of Au$_{147}$ clusters with different initial structures]. See DOI: 00.0000/00000000.}



\section{Introduction}
Gold nanoclusters have been studied extensively compared to the other metallic systems owing to their fundamental and technological importance. Despite being inert in the bulk state, gold exhibits high catalytic activity at the nanoscale,\cite{haruta1998AuCat,whetten2000AuCat,rev2005AuCat,rev2006AuCat,chen2009AuCat,kawasaki2012AuCat,rev2018AuCatAPCs} 
which is strongly influenced by the size and shape of the nanoclusters. For instance, it was shown that Au particles with sizes greater than 2 nm are completely inactive\cite{turner2008AuCatSizeEffect} for styrene epoxidation by O$_2$ and this sharp size dependence was attributed to the electronic structure of small Au nanoclusters. In addition, the various geometrical shapes (such as icosahedron, decahedron and truncated octahedron) have different number of low coordination sites\cite{li2015AuCatShapeEffect} (vertex and edge sites) which influences the catalytic activity. Along with catalysis, Au particles have been explored for various applications such as biosensors,\cite{rev2004AuBioSens} plasmonics,\cite{rev2005AuSPR,rev2017AuSPR} \emph{in vivo} and biomedical applications.\cite{colleen2019AuinVivoDisease}$^,$\cite{rev2020AuBioMedical} Given the strong structure-property relation, numerous theoretical studies have focused on determining the lowest energy structure of Au nanoclusters.

Gold exhibits strong relativistic effects\cite{pyykko2004AuRelEffects} which results in significant \emph{sd} hybridization and contraction of the \emph{6s} orbital. The nature of Au-Au interactions are short range and the bonds are anisotropic.\cite{garzon1998AuRelSRAmor}$^,$\cite{apra2006AuRelSRAmor} Due to these properties, Au exhibits very interesting structures which differ from other metal systems that do not have strong relativistic effects. At very small sizes, the Au nanoclusters adopt planar structures\cite{hakkinen2000AuPlanarStructs,perez1999AuPlanar} and depending on the charge of the clusters, the transition from planar 2D to 3D structures occurs around the size of 13, 14 atoms.\cite{hakkinen2003AuPlanarStructs} In contrast, the 2D to 3D transition for Cu and Ag (which are in the same group) occurs at much lower sizes.\cite{hakkinen2002Au2Dto3DCuAg} The planar structures have been experimentally confirmed\cite{furche2002AuPlanarStructsExpIon},\cite{hakkinen2003AuPlanarStructs} through ion mobility measurements or photoelectron spectroscopy combined with density functional calculations. There is a transition to hollow cage-like structures for Au nanoclusters containing more than 14 atoms up to a size of ca. 35 atoms.\cite{johansson2004AuCageStructs}$^-$\nocite{gu2004AuCageStructs}\nocite{fa2006AuCageStructs}\cite{xing2006AuCageStructs} A highly symmetric hollow tetrahedron has been predicted as the global minimum structure of Au$_{20}$ cluster through an unbiased search.\cite{apra2006AuRelSRAmor} The planar and cage-like structures are also the lowest energy structures for Au nanoclusters supported on flat MgO\cite{riccardo2009AuFlatMgO}$^,$\cite{riccardo2011AuFlatMgO} and stepped MgO\cite{damianos2012AuSteppedMgO} surfaces in the size range of 11 to 40 atoms.

At relatively larger sizes, Au nanoclusters adopt compact structures. Garzón \emph{et al.}\cite{garzon1998AuRelSRAmor,michaelian1999AuRelSRAmor} carried out global optimization using a version of the Gupta potential\cite{gupta1981}$^,$\cite{rgl1989} for Au$_n$ clusters with n = 38, 55, and 75. It was found that, at the sizes of 38 and 55, the lowest energy configurations have an amorphous-like structure. However, at the size of 75, decahedron is the lowest energy structure. This was also confirmed at the DFT level under the local density approximation (LDA). Although, 55 is a ``magic'' size corresponding to icosahedron, Au nanoclusters do not prefer the icosahedral arrangement. Huang \emph{et al.}\cite{huang2008AuExpPESDFTnonIco} studied the structure of Au anion clusters in the size range of 55 to 64 atoms through a combination of photoelectron spectroscopy of density functional calculations. They observed that Au$_{55}^-$ indeed adopts a non-icosahedral structure with very low symmetry. Bao \emph{et al.}\cite{bao2009AuBHMC} calculated the global minimum structures for Au clusters up to a large size of 318 atoms. At the sizes corresponding to a perfect icosahedron (55, 147, and 309), icosahedron is not the global minimum and \emph{fcc} structure is found to be the global minimum for Au$_{55}$.\cite{bao2009AuBHMC} However, recently, Schebarchov \emph{et al.}\cite{schebarchov2018AuHSA} showed that, although the \emph{fcc} structure is the global minimum of Au$_{55}$ according to the Gupta potential, disordered (which is referred to as ``ambiguous'') and twinned \emph{fcc} structures become the lowest energy configurations (both are degenerate) at the DFT level.

Nelli \emph{et al.}\cite{nelli2020AuBHMC} carried out global optimization of Au$_{147}$ and Au$_{294}$ clusters using a Gupta potential. They found that global minimum structure of Au$_{147}$ is a Marks decahedron which is formed by adding a single atom on the \{100\} facet of the perfect 146-atom Marks decahedron. This structure has been confirmed to be the global minimum at the DFT level as well.\cite{palomaresBaez2017AuDFT} Marks decahedron is also the global minimum for Au$_{294}$. From the global optimization studies, it is clear that Au clusters disfavor the icosahedral structure. This is also true at the geometrical ``magic'' sizes corresponding to perfect icosahedra.

Electron microscopy imaging is a powerful technique to study the structure of Au particles experimentally. Wang \emph{et al.}\cite{wang2012AuImaging} studied the structural transitions of Au$_{923}$ clusters under the influence of an electron beam. They observed that icosahedral clusters transformed to either decahedral or fcc structures. On the other hand, decahedral and fcc structures remained stable under the electron beam. Young \emph{et al.}\cite{young2010AuImagingInSitu} carried out \emph{in situ} heating and cooling under an electron microscope. Particles with initial structure of \emph{fcc} (5.5 nm), icosahedron (7.2 nm), and decahedron (10.4 nm) all transformed to a final decahedral structure on heating indicating a significant stability of the decahedral motif in this size range (2 nm to 15 nm). From an application viewpoint, it is essential to understand the structural distribution of Au particles as a function of temperature; this is particularly urgent for catalysis where the catalysts are expected to function at elevated temperatures.

Wells \emph{et al.}\cite{wells2015AuImagingACFraction} calculated the fraction of as-deposited Au clusters at the sizes of 561, 742, and 923 using HAADF STEM images. At all the sizes, decahedron is the abundant motif (41 to 46\%) followed by \emph{fcc} structures (31 to 37\%) along with less than 5\% icosahedra. In another study, Foster \emph{et al.}\cite{foster2018AuImagingACFraction} estimated the fraction of various structural motifs (\emph{fcc}, decahedron, icosahedron) of Au$_{561}$ as a function of temperature (20 $^\circ$C to 500 $^\circ$C). At temperatures greater than 125 $^\circ$C equilibration was obtained, with \emph{fcc} structures being more likely than decahedral ones, while icosahedra were absent. This is completely different to the case of as-deposited Au$_{561}$ where decahedral structures were found to be more likely than fcc ones due to kinetic trapping.\cite{wells2015AuImagingACFraction} In both the studies, there is a significant fraction of the structures which have been classified as unidentified/amorphous owing to the inherent difficulty in characterizing the structures from 2D electron microscopy images. Although, the structural distribution can be obtained from experimental studies, it is far from complete. Hence, a theoretical method which can provide a reference for the temperature-dependent structural distribution is necessary. Unfortunately, each method available at present suffers from some type of limitation that restricts the temperature range of applicability. 

A well known approach is the Harmonic Superposition Approximation (HSA) to the partition function.\cite{hsa1993}  In this method, many low energy minima ($\sim 10^4$) are collected and the partition function is constructed which is then used to calculate the probability of a structure as a function of temperature. HSA has been used to study the solid-solid transitions and calculate the occurrence probability of various geometrical motifs in Lennard Jones (LJ) clusters\cite{doye2001hsaLJ}$^-$\nocite{doye2002hsaLJ}\nocite{mandelshtam2006hsaLJ}\cite{sharapov2007hsaLJ}, Au nanoclusters\cite{schebarchov2018AuHSA}, Pd-Pt nanoalloys,\cite{panizon2015hsaPdPt} and Ag-Cu, Ag-Ni nanoalloys.\cite{bonventre2018hsaAgCuANi} HSA is in principle exact in the limit $T\to0$, since it can easily incorporate quantum corrections, and it is in general very accurate at low temperatures, which may extend  well above room temperature for noble and transition metals. HSA becomes progressively erroneous as the temperature gets close to the melting range as the anharmonic effects are not accounted for.\cite{rossi2004pIh} \rev{Anharmonic effects were experimentally observed\cite{bracco1996Ag110AnharmonicRT} well below the melting point at the onset of roughening transition of Ag (110) surfaces.} But this is not the only problem of HSA at high temperatures.
In fact, when temperature approaches the melting range, an increasingly large number of high-energy liquidlike local minima must be taken into account. Collecting these local minima becomes very cumbersome and not always straightforward. 

Another widely used technique is conventional molecular dynamics (MD). MD performs poorly at low temperatures in which the free-energy barriers are much larger than the thermal energy, leading to inadequate sampling of phase space. \cite{Bonella2012epjb,panizon2015hsaPdPt} Typically, for noble and transition metal clusters, one may achieve a good sampling of configuration space only at temperatures close to the melting range, where transition between relevant structural motifs become possible within the limited time scale of MD.

In order to overcome the limitations of poor sampling, one can in principle momentarily heat up the system, in order to promote thermally-activated barrier crossing. In order to recover the correct behavior at the target temperature, multiple configurations or \emph{replicas} of the system are used which are allowed to exchange configurations at fixed intervals according to a Metropolis-like criterion. The replicas are at different temperatures with the high temperatures allowing to sample effectively phase space. Such methods are generally referred to as \emph{replica exchange} or \emph{parallel tempering}.\cite{pearl2005PT} Parallel tempering can be combined with either Monte Carlo or molecular dynamics. Parallel Tempering Monte Carlo (PTMC) has been employed to study the structural transitions and the structural distributions in Lennard Jones clusters,\cite{neirotti2000PTMCnLJ}$^-$\nocite{calvo2000PTMCnLJ}\nocite{noya2006PTMCnLJ}\cite{cezar2017PTMCnLJ} Argon clusters,\cite{senn2014PTMCnAr} metal \& nanoalloy clusters.\cite{calvo2008PTMCnPdPt}$^-$\nocite{calvo2008PTMCnAgM}\nocite{calvo2011PTMCnCoPt}\cite{cezar2019PTMCnPtCoNi} In PTMC simulations, typically, random displacement moves are used which could hinder the inter-motif transition at larger sizes. For instance, Pd-Pt nanoalloys (sizes 55, 147, and 309) retained the initial icosahedral geometry during the PTMC simulations,\cite{calvo2008PTMCnPdPt} although it is expected that several geometrical motifs co-exist at any combination of size, composition, and temperature. In contrast to PTMC, Parallel Tempering Molecular Dynamics (PTMD) studies are less common for clusters. Using PTMD, size-dependent melting of Fe particles\cite{calvo2012PTMDnFeMelting} and low energy structures of Au clusters have been studied.\cite{tarrat2018PTMDnAu} 

PTMD offers a few advantages over the PTMC method. Compared to PTMC, PTMD is relatively straightforward in that one does not have to search for the structures as required in PTMC. Additionally, molecular dynamics at high temperatures can avoid the issue of being trapped in a specific geometrical motif. However, even the PTMD sampling may become cumbersome at low temperatures, that typically means below room temperature for noble and transition metals. In addition, PTMD is a purely classical method that does not incorporate quantum corrections to the occupation probabilities of the different minima. These corrections may be important at low temperatures.\cite{panizon2015hsaPdPt}

In the current work, we propose a simulation methodology that combines different techniques in order to obtain an accurate sampling of the configuration space of nanoclusters in the full temperature range from 0 K up to melting. The methodology is mainly based on PTMD simulations complemented by global optimization searches and HSA calculations. It consists of three steps
\begin{itemize}
    \item[1.] Global optimization searches (here performed by the Basin Hopping algorithm \cite{Wales1997jpca}) in order to search for the lowest-energy structures in the energy landscape of the nanoclusters. These will be part of the input of the PTMD simulations and of the HSA calculations
    \item[2.] PTMD simulations in which the convergence is checked also against the choice of the initial configurations. In the PTMD simulations, configurations are frequently sampled to produce a collection of local minima to be used in the HSA calculations
    \item[3.] HSA calculations that make use of the local minima collected in step 2., complemented by minima from step 1. if necessary.
\end{itemize}
\rev{Accuracy of HSA decreases with increasing temperature. In PTMD simulations, exchange of configurations between replicas leads to an improved sampling. As a result of these two factors, PTMD will give an accurate sampling of the configuration space at higher temperatures. Typically, there will be an intermediate temperature range (which may change from system to system) in which the results of HSA calculations and of PTMD simulations are in good agreement with each other. For temperatures below this range, HSA results can be safely trusted.} \rev{We note that in principle the manner in which we combine the PTMD and HSA is not an on-the-fly combination. A previous study\cite{sharapov2007combinePTMDnHSA} has indeed combined HSA and Parallel Tempering Monte Carlo (PTMC) by treating an auxiliary HSA system as an additional replica. Similarly, HSA has been combined with ``basin-sampling''\cite{bogdan2006hsaBasinSampling} to improve the accuracy. HSA has also been employed to identify the optimal temperatures\cite{ballard2014superposOptTemp} of the replicas during PTMC which allows to maintain uniform acceptance rates.}

We use this method to study the temperature-dependent structures of Au nanoclusters. We consider three cluster sizes: Au$_{90}$, Au$_{147}$, and Au$_{201}$ in the size range of 1 $-$ 2 nm where the clusters are expected to exhibit high catalytic activity. 147 and 201 are geometric ``magic'' sizes corresponding to icosahedron and truncated octahedron, respectively. Au$_{147}$ has been studied previously using HSA\cite{schebarchov2018AuHSA} which allows for a comparison with our results. On the other hand, Au$_{90}$ is not a geometric magic size for any motif.

Our calculations will give accurate descriptions of the competition between structural motifs for $T$ from 0 K to the complete melting. Different types of structural transitions at the solid state will be singled out.

\section{Methods}

\subsection{Interatomic Potential}

We model the interaction between the Au atoms using a tight binding model within the second moment approximation (TBSMA)\cite{tbsma1971} which is also referred to as Gupta\cite{gupta1981} potential or Rosato-Guillope-Legrand (RGL)\cite{rgl1989} potential. In this model, the cohesive energy of a cluster is defined as
\begin{equation} \label{eqnCohEnergyGupta}
    E_{c}= \sum_{i}\Bigg\{\sum_{j}Ae^{-p(r_{ij}/r_0-1)}-\sqrt{\sum_{j}\xi^2e^{-2q(r_{ij}/r_0-1)}} \Bigg\}
\end{equation}
where $r_0$ is the bulk nearest neighbor distance. $r_{ij}$ is the distance between the atoms $i$ and $j$. $A$, $\xi$, $p$, and $q$ are the model parameters. The potential is represented by a continuous $5^{th}$ degree polynomial beginning from the second nearest neighbor distance which goes to zero at the third nearest neighbor distance. The parameters used in this work are given in \citet{baletto2002potParams}. This potential has been used to study the structure of Au nanoclusters during the gas-phase growth and on MgO substrates obtaining very good agreement with the  experiments.\cite{wells2015AuImagingACFraction}$^,$\cite{han2014imagingOnMgO} Additionally, the structural predictions of the potential relating to the surface defects of the ``rosette'' type\cite{apra2004AuRosette} and the propensity to disfavor icosahedral motif agree with the DFT calculations.\cite{palomaresBaez2017AuDFT} With respect to the bulk systems, the potential correctly predicts the reconstruction of the \{110\} surfaces\cite{guillope1989surfRecons} and the activation barrier of self diffusion on \{111\} surfaces (0.12 eV)\cite{ferrando1995surfDiffusion} which is confirmed by DFT calculations.\cite{boisvert1995surfDiffusionDFT}

\subsection{Global optimization}
Before PTMD simulations, we carry out global optimization searches in order to understand the global minimum and the low energy structures for each cluster size. These structures serve as the initial structures for PTMD simulations. We employ Monte Carlo combined with basin hopping referred to as the basin hopping Monte Carlo (BHMC)\cite{nelli2020AuBHMC}$^,$\cite{rossi2009BHMC} for searching the low energy structures. We use a highly disordered structure to initialize the BHMC search. At each size, we run at least 5 independent searches of at least $2.5\times 10^5$ steps.

\subsection{Parallel Tempering Molecular Dynamics (PTMD)}
In conventional molecular dynamics, when  the free-energy barriers separating different structures are larger than the thermal energy $k_B T$ available to the system, with $k_B$ the Boltzmann constant, the simulations can get trapped in local free-energy minima for long times, thus preventing an effective sampling of all the representative configurations at these temperatures. In simulations of nanoclusters, this issue is sometimes overcome by raising the temperature close to the melting one.\cite{Nelli2019nanoscale}  
In PTMD, this strategy is further extended by simulating different copies (replicas) of the system at temperatures ranging from low to high; from time to time, exchange of configurations between the replicas is attempted according to a Metropolis-like criterion. The replicas at higher temperatures are able to overcome free-energy barriers and thus sample a larger number of configurations compared to the replicas at lower temperatures. Due to the exchange of configurations between the replicas (``tempering''), the sampling of the phase space improves significantly for all temperatures. 

In our PTMD simulation, we consider $M$ replicas and each replica is thermostatted at temperature T$_m$ $(m=1,2,...,M)$ in a canonical ensemble. For the replica exchange, the set of nearest neighbor pairs, $\mathcal{N}{\equiv}\{(m,m+1)|\,m=1,2,...,M-1\}$, is divided in to two subsets: even pairs, $\mathcal{E}{\equiv}\{(2j,2j+1)\}$ and odd pairs, $\mathcal{O}{\equiv}\{(2j-1,2j)\}$. Depending on how the exchange pairs are selected, there are two commonly used algorithms.\cite{lingenheil2009DEOSEO} In the deterministic even/odd (DEO) algorithm, the exchange attempts alternate between even and odd pairs. In the case of stochastic even/odd (SEO) algorithm, either an even or an odd pair is chosen randomly. In this work, we use the stochastic even/odd (SEO) algorithm for replica exchange. An exchange attempt between the adjacent replicas \emph{p} and \emph{q} is either accepted or rejected according to a probability given by the following Metropolis-like criterion.
\begin{equation}
    p=min\{1,e^{-(E_{q}-E_{p})(\beta_{q}-\beta_{p})}\}
\end{equation}
where $\beta_{p}=1/k_{B}T_p$, $\beta_{q}=1/k_{B}T_q$. The potential energies of the replicas $p$, $q$ are $E_p$, $E_q$ respectively.

A key step in PTMD simulations is the choice of the number of replicas and the associated temperatures.\cite{pearl2005PT} We carried out a preliminary PTMD simulation to identify the melting range of the clusters at all the three sizes of 90, 147, and 201. Based on the melting region and given that we are interested in studying the finite-temperature structures, we chose the temperature ranges of 250 K $-$ 550 K (Au$_{90}$), 300 K $-$ 600 K (Au$_{147}$), and 300 K $-$ 650 K (Au$_{201}$). The number of replicas, $M$ ranges from 24$-$36 depending on the size of the cluster; the criterion used is to have at least 20 to 30 \% acceptance rate for replica exchanges. The temperatures follow a geometric progression\cite{pearl2005PT,cezar2017PTMCnLJ} with additional replicas close to the melting region to ensure finer temperature intervals. All the simulations have been carried out using LAMMPS\cite{lammps} package. 

We use a time step of 5 fs. Exchanges are attempted every 250 ps and they are accepted or rejected according to the Metropolis-like criterion given above. For the structural analysis, we consider the configurations at 125 ps after every exchange attempt. 
In order to ensure we achieved convergence of the structural distribution in PTMD, we adopt the following procedure. We begin the simulations with all the replicas having the same initial configuration. During the initial phase of the simulations, the structures retain the memory of the initial configuration until a sufficient number of successful exchanges is accomplished. Hence, we disregard the initial phase, which ranges from 0.5 $\mu$s to 1.5 $\mu$s, and start collecting statistics of the configurations after a certain time, for at least 2 $\mu$s. We have verified that this procedure results in convergence by starting from several different initial configurations, including random clusters, global minima, and a mixture of representative structures, see Fig.~\ref{fgr:initialStructsAu147} and the Supplementary Information. 

Additionally, we performed very long conventional MD simulation of Au$_{201}$ at $300$~K, to demonstrate that the replica exchange indeed improves the sampling in the present case. These simulations start from the truncated octahedron, which, as we will see in the following, is the lowest-energy structure for size 201 but is very unlikely to be observed at 300 K, where decahedra dominate due to entropic effects. In more than 60 $\mu$s of conventional constant temperature MD, no transitions from the initial truncated octahedron to other structures were observed, showing that the sampling given by MD at 300 K is largely incomplete.

This result, together with the check on the independence on initial conditions, provides a strong evidence for practitioners that the PTMD approach is effective at sampling the structure distribution even at temperatures lower than the melting one and that is rather insensitive to the choice of initial structures.

\begin{table}[!b]
\small
\caption{\rev{Number of local minima used for HSA analysis. At each size, the minima were collected up to an energy (E\textsubscript{cutoff}) above the global minimum. E\textsubscript{cutoff} values at the sizes 90, 147, and 201 are 1.0 eV, 1.5 eV, and 1.0 eV respectively.}}
\centering
{\def\arraystretch{1.15}
\begin{tabular*}{0.45\textwidth}{@{\extracolsep{\fill}}l|lll}
\hline
\multirow{2}{2em}{\rev{Structure}} & \multicolumn{3}{c}{\rev{Size}} \\
& {\rev{90}} & {\rev{147}} & {\rev{201}}\\
\hline
{\rev{amorphous}} & {\rev{69}} & {\rev{$-$}} & {\rev{$-$}} \\
{\rev{fcc}} & {\rev{316}} & {\rev{1382}} & {\rev{728}} \\
{\rev{twin}} & {\rev{3107}} & {\rev{3651}} & {\rev{1316}} \\
{\rev{icosahedron}} & {\rev{1}} & {\rev{2444}} & {\rev{$-$}} \\
{\rev{decahedron}} & {\rev{994}} & {\rev{8554}} & {\rev{8583}} \\
{\rev{mix}} & {\rev{6632}} & {\rev{1919}} & {\rev{$-$}} \\
\hline
{\rev{all}} & {\rev{11119}} & {\rev{17950}} & {\rev{10627}} \\
\hline
\end{tabular*}
}
\label{tab:HSAconfigs}
\end{table}

\begin{table*}[!t]
\small
\caption{CNA signatures of the various coordination types considered for the structure classification}
\centering
{\def\arraystretch{1.15}
\begin{tabular*}{0.75\textwidth}{@{\extracolsep{\fill}}lc}
\hline
{Coordination type} & {CNA signature} \\
\hline
inner-fcc & (421) (421) (421) (421) (421) (421) (421) (421) (421) (421) (421) (421) \\
inner-hcp & (421) (421) (421) (421) (421) (421) (422) (422) (422) (422) (422) (422) \\
inner-5-axis & (422) (422) (422) (422) (422) (422) (422) (422) (422) (422) (555) (555) \\
central-Ih & (555) (555) (555) (555) (555) (555) (555) (555) (555) (555) (555) (555) \\
\hline
\end{tabular*}
}
\label{tab:cnaSignatures}
\end{table*}
\begin{table*}[!t]
\small
\caption{Rules for classification of the structures. The values referring to the amount of a given coordination type are in percentage}
\centering
{\def\arraystretch{1.15}
\begin{tabular*}{0.75\textwidth}{@{\extracolsep{\fill}}ll}
\hline
{Structure} & {Classification rule}\\
\hline
amorphous & non-classified > 60\\
fcc & inner-5-axis = 0 and inner-fcc > 10 and inner-hcp = 0\\
twin & inner-5-axis = 0 and (inner-hcp + inner-fcc) > 10 and inner-hcp > 0\\
icosahedron & central-Ih > 0\\
decahedron & inner-5-axis > 0 and (inner-hcp + inner-fcc) > 20\\
mix & rest\\
\hline
\end{tabular*}
}
\label{tab:classRules}
\end{table*}

\subsection{Harmonic Superposition Approximation}
The application of the Harmonic Superposition Approximation (HSA) to nanoclusters has been discussed in numerous works.\cite{schebarchov2018AuHSA,panizon2015hsaPdPt,doye2002hsaLJ,bonventre2018hsaAgCuANi} Here, we only provide the relevant equations. In HSA method, the partition function is written as,
\begin{equation} \label{eqnPartFunc}
    Z=\sum_{i}\frac{e^{-\beta{E^0_i}}Z^{tr}_iZ^{rot}_iZ^{vib}_i}{g_i}
\end{equation}
where $\beta=1/(k_BT)$. The summation is over all the local minima, $i$, considered for the HSA. $E^0_i$ is the energy of the local minimum, $i$. $Z^{tr}$, $Z^{rot}$,  and $Z^{vib}$ are the translational, rotational, and vibrational contributions to the partition function, respectively. The translational contribution is equal for all clusters with same number of atoms. It has been shown that the rotational contribution does not affect the probability of the local minima.\cite{panizon2015hsaPdPt} Hence, we consider only the vibrational contribution. The denominator, $g_i$, is the order of the symmetry group of the local minimum $i$. The vibrational contribution due to a single minimum is given by
\begin{equation} \label{eqnPartFuncVib}
    Z^{vib}=\prod_{n=1}^{3N-6}\frac{e^{-{\beta}{\hbar}{\omega}_n/2}}{1-e^{-{\beta}{\hbar}{\omega}_n/2}} \text{ ,}
\end{equation}
where $\omega_n$ are the $3N-6$ ($N$ is the number of atoms in the cluster) frequencies of the normal modes. The probability of a local minimum as a function of temperature is now given by
\begin{equation} \label{eqnProbHSA}
    p_i=\frac{e^{-\beta{E^0_i}}Z^{vib}_i}{\sum_{j}e^{-\beta{E^0_j}}Z^{vib}_j} \text{ .}
\end{equation}
We define the probability of a specific structure type ($p^{struct}$) by summing up the probabilities of all the minima belonging to that structure type.
\begin{equation} \label{eqnProbHSAStruct}
    p^{struct}=\sum_{k}p_k
\end{equation}
where $k$ represents all the minima having the same structure. The set of local minima used for the HSA analysis have been collected from the PTMD simulations up to an energy of about 1.0$-$1.5 eV above the global  energy minimum. \rev{Two minima were considered to be distinct if they had different structure and their energies were separated by at least 0.05 meV for all the sizes. The number of local minima considered for the HSA analysis at each size is reported in Table \ref{tab:HSAconfigs}.}

\subsection{Structure classification}

\begin{figure}[!b]
\centering
  \includegraphics[width=7cm]{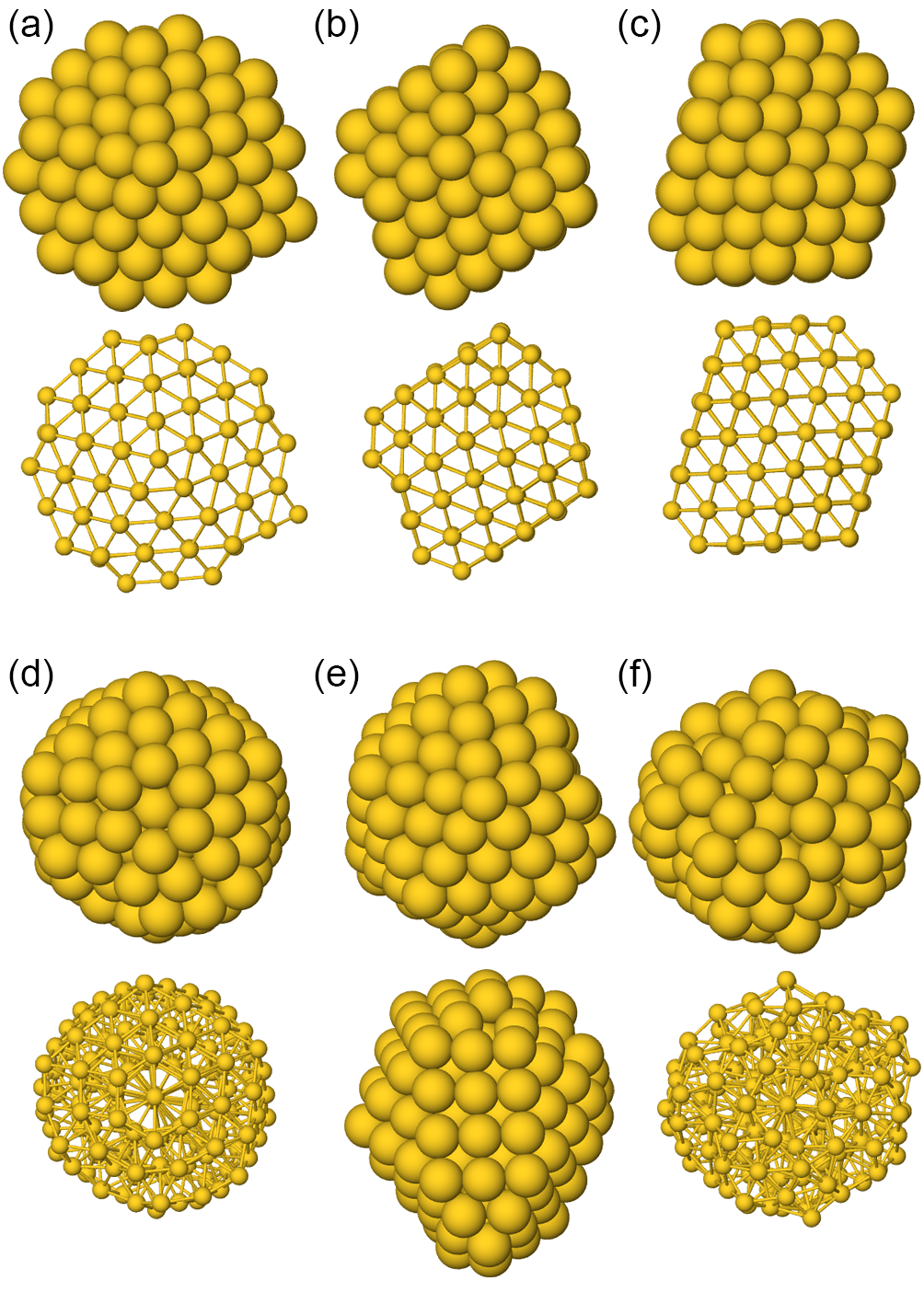}
  \caption{Representative structures of Au$_{147}$ clusters. (a) decahedron (Dh), (b) twin, (c) fcc, (d) icosahedron (Ih), (e) an example of mix structure having the features of both icosahedron and decahedron, and (f) amorphous.}
  \label{fgr:representativeStructures}
\end{figure}
In order to calculate the fraction (PTMD simulations) or the probability  (HSA) of the various structures as a function of temperature, a structure classification method is required. Here, we employed the Common Neighbor Analysis (CNA)\cite{cna1994} method. In CNA, each pair of atoms is assigned a set of three integers (r,s,t) which gives an indication of the coordination of these atoms. For example, (421) and (422) indicate \emph{fcc} (face centered cubic) and \emph{hcp} (hexagonal close packed) coordination, respectively. In any cluster, the coordination environment of each atom can be defined by the CNA signature associated with all the nearest neighbors. For example, an inner atom in a truncated octahedron has 12 nearest neighbors and each nearest neighbor has (421) coordination. Similarly, a surface atom on the \{111\} facet has 9 neighbors (6 on the surface and 3 inner neighbors). The CNA signature for this atom is \{(311),(311),(311),(311),(311),(311),(421),(421),(421)\}. Based on the CNA signature associated with the nearest neighbors, each atom can be assigned a coordination type. 

In this work, we consider five coordination types for classifying the structure of a cluster: \emph{inner-fcc}, \emph{inner-hcp}, \emph{inner-5-axis}, \emph{central-Ih}, and \emph{non-classified}. The CNA signatures of these coordination types are given in Table \ref{tab:cnaSignatures}. The \emph{inner-fcc} and \emph{inner-hcp} atoms have \emph{fcc} and \emph{hcp} coordination. Atoms having \emph{inner-5-axis} coordination belong to a local 5-fold axis. The \emph{central-Ih} coordination indicates an atom that is surrounded by 12 neighbors each associated with a (555) signature. This is typical of icosahedral structures. Atoms having \emph{non-classified} coordination do not have a well defined coordination.

We consider six different classes of structure depending on the percentage of various coordination types in a given cluster: decahedron (Dh), twin, fcc (truncated octahedron), icosahedron (Ih), mix, and amorphous. We define the following simple rules for structure classification (representative structures are shown in Fig.~\ref{fgr:representativeStructures}). All clusters that have at least 60\% \emph{non-classified} coordinated atoms are classified as \emph{amorphous} structures. Next we consider the \emph{inner-5-axis} coordination. Clusters that have non-zero \emph{inner-5-axis} coordination, can have either icosahedral or decahedral structure. Icosahedral clusters have a \emph{central-Ih} coordinated atom. Clusters than do not have a \emph{central-Ih} coordinated atom but have at least 20\% combined \emph{inner-fcc} and \emph{inner-hcp} are classified as decahedra. Clusters with \emph{twin} and \emph{fcc} structures have at least 10\% combined \emph{inner-fcc}, \emph{inner-hcp} atoms, and zero \emph{inner-5-axis} atoms. Clusters with \emph{twin} structure have non-zero \emph{inner-hcp} atoms. All other clusters are classified as \emph{mix} structures. All the rules are listed in Table \ref{tab:classRules}. The \emph{mix} structures have features of more than one geometrical motif and are discussed in detail in Sec.~\ref{sec:mixed}. An example is shown in Fig. \ref{fgr:representativeStructures}e where the features of icosahedron and decahedron are observed in the same cluster. We treat this structure type in greater detail and will be discussed later.

\begin{figure}[!b]
\centering
  \includegraphics[width=7.5cm]{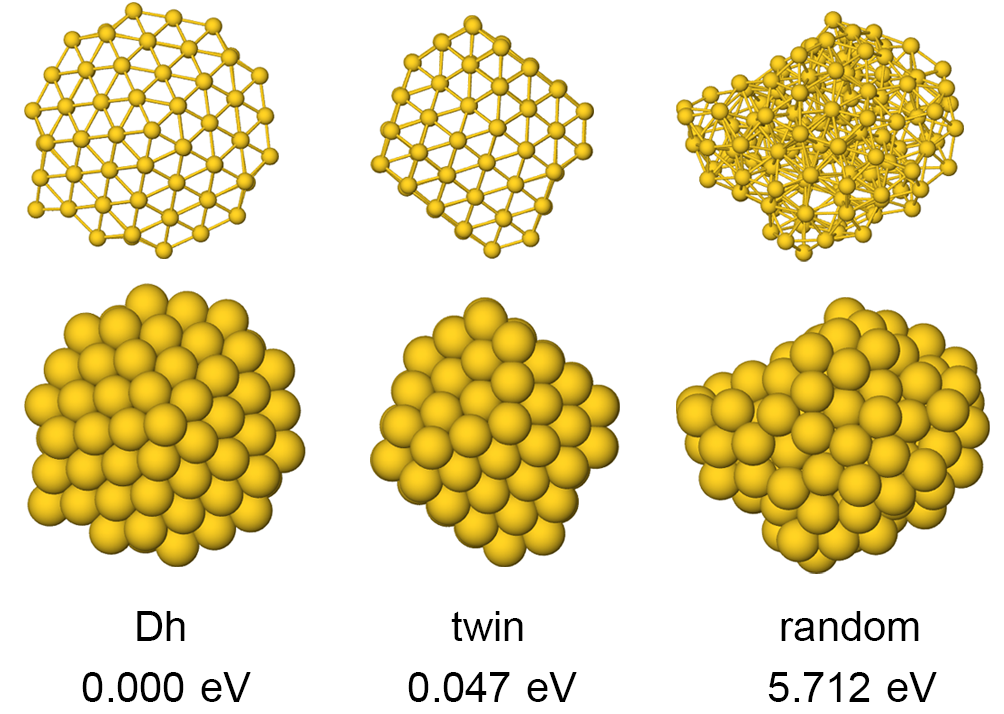}
  \caption{The initial structures considered for PTMD simulations of Au$_{147}$ clusters along with their energy. The energy is measured relative to the global minimum.}
  \label{fgr:initialStructsAu147}
\end{figure}

\section{Results and Discussion}
\rev{In order to obtain the melting point, we calculated the heat capacity C\textsubscript{V} from the PTMD simulations. The melting point was estimated as the temperature corresponding to the peak value of C\textsubscript{V}.} The melting points of Au$_{90}$, Au$_{147}$, and Au$_{201}$ calculated from PTMD simulations are 420 K, 505 K, and 550 K, respectively. In general, Gupta potential underestimates the melting point.\cite{guptaPotUnderEstMP2006} However, the structural predictions of the Gupta potential used in the current work have very good agreement with experiments and DFT calculations.\cite{palomaresBaez2017AuDFT,wells2015AuImagingACFraction,han2014imagingOnMgO,apra2004AuRosette} As a result, we believe that the PTMD results will be qualitatively accurate at the very least. In the following, we analyze in detail the results for each cluster size, starting from Au$_{147}$ for which a detailed HSA analysis was available.\cite{schebarchov2018AuHSA}

\subsection{Au$_{147}$}
Our BHMC searches identify the same global minimum -- decahedron, shown in Fig. \ref{fgr:initialStructsAu147} -- which has been reported in previous studies.\cite{schebarchov2018AuHSA,nelli2020AuBHMC,palomaresBaez2017AuDFT} We find a twin (Fig. \ref{fgr:initialStructsAu147}) which is about 50 meV higher than the global minimum energy. The best fcc and icosahedron are \rev{0.109} and 0.910 eV higher in energy than the global minimum, respectively. The best icosahedron has a defective shell with three surface ``rosette'' defects. We find that there are several decahedra which compete closely with the global minimum and lie within 100 meV.\cite{nelli2020AuBHMC} These results show that icosahedra are not favored energetically at 0 K. In the following, we verify whether this trend persists at finite temperatures as well.

\begin{figure}[!t]
\centering
  \includegraphics[width=8.5cm]{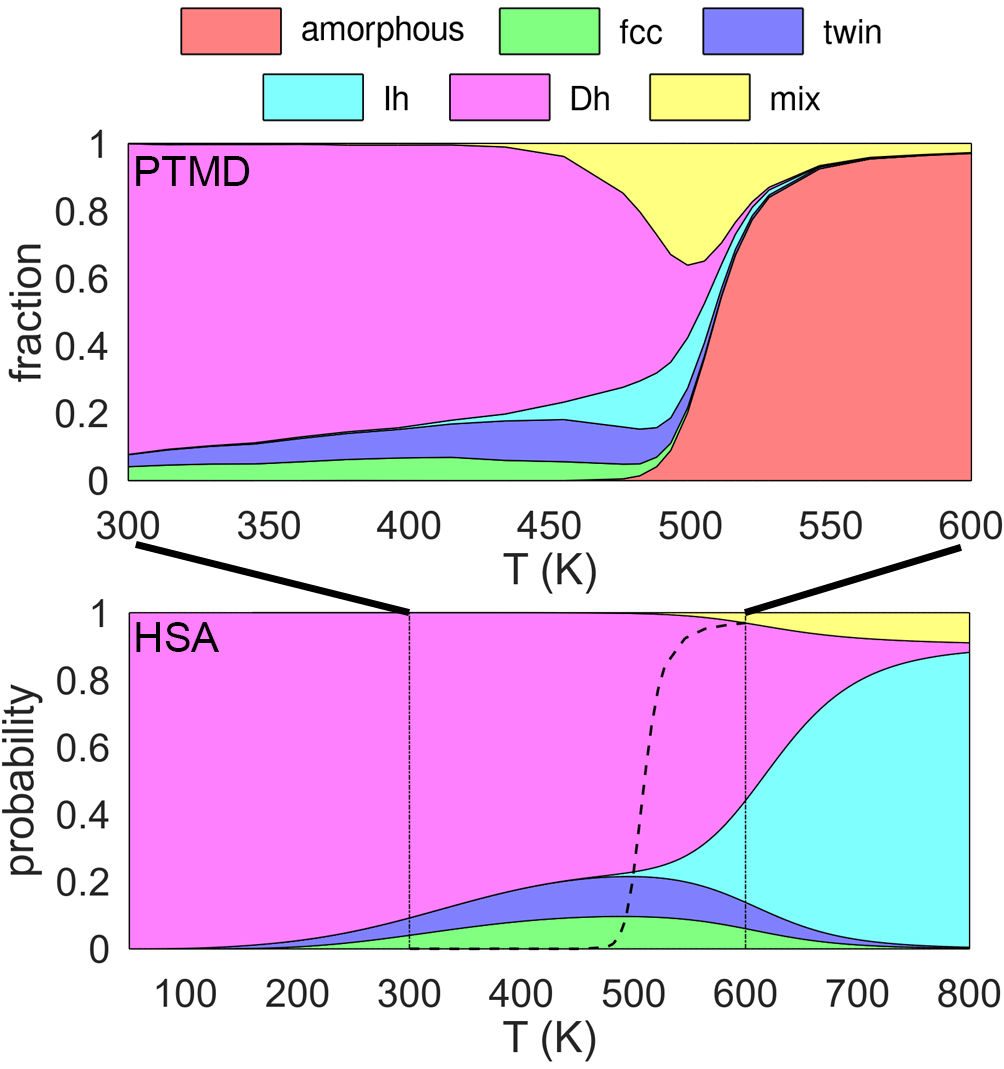}
  \caption{(top) Occurrence fraction of various structures of Au$_{147}$ nanoclusters as a function of temperature in the range 300 K $-$ 600 K obtained via PTMD simulations. (bottom) Probability of various structures of Au$_{147}$ nanoclusters as a function of temperature in the range 50 K $-$ 800 K obtained via HSA. For comparison, we report with vertical lines  the range of PTMD temperatures and with a dashed line the fraction of amorphous structures calculated from PTMD simulations.}
  \label{fgr:ptmdAu147}
\end{figure}

\begin{table*}[!t]
\small
\caption{Comparison of the occurrence fraction from PTMD and the probability from HSA of various structures at selected temperatures. The values in brackets are the probabilities calculated using HSA. Amorphous structures were not considered for the HSA analysis due to their high energy.}
\centering
{\def\arraystretch{1.15}
\begin{tabular*}{1.00\textwidth}{@{\extracolsep{\fill}}lllllll}
\hline
{Structure} & {300 K} & {345 K} & {396 K} & {455 K} & {499 K} & {511 K} \\
\hline
{amorphous} & {0.02 (NA)}& {0.00 (NA)}& {0.02 (NA)}& {0.01 (NA)}& {20.18 (NA)} & {54.36 (NA)} \\
{fcc} & {4.12 (4.03)}& {4.95 (6.08)}& {6.70 (8.06)}& {5.59 (9.41)}& {1.40 (9.61)} & {0.34 (9.52)} \\
{twin} & {3.55 (5.18)}& {5.93 (7.25)}& {8.46 (9.53)}& {12.51 (11.39)}& {5.82 (11.91)} & {2.15 (11.87)} \\
{Ih} & {0.10 (0.00)}& {0.34 (0.00)}& {0.49 (0.00)}& {5.18 (0.17)}& {14.97 (1.33)} & {7.18 (2.17)} \\
{Dh} & {92.12 (90.80)}& {88.44 (86.67)}& {83.86 (82.40)}& {72.86 (78.99)}& {21.53 (76.95)} & {6.42 (76.13)} \\
{mix} & {0.10 (0.00)}& {0.34 (0.00)}& {0.48 (0.00)}& {3.85 (0.04)}& {36.11 (0.20)} & {29.56 (0.30)} \\
\hline
\end{tabular*}
}
\label{tab:fracPTMDprobHSA}
\end{table*}

We considered three distinct initial structures for the PTMD simulations: the global minimum decahedron, twin, and a completely nonphysical random structure (Fig. \ref{fgr:initialStructsAu147}). Early in the simulations ($\sim$ 0.5$-$0.6 $\mu$s), there are few differences (see Fig.~S1). The fraction of twin is higher when we begin with a twin initial structure while it is lower when the initial structure is a Dh. On the other hand, we observe a negligible amount of fcc when we begin with a random structure. However, these differences decrease as simulations run for a longer time; we find that all three simulation converge to the same structural distribution after 2.4 $\mu$s. This shows that the PTMD simulations are not sensitive to the initial structure given that they are run for ``sufficiently'' long time.

The finite-temperature structures of Au$_{147}$ obtained from PTMD simulations are shown in Fig.~\ref{fgr:ptmdAu147}. The fraction of structures as a function of temperature is calculated as the mean of the three simulations with different initial structures (Dh, twin, and random). The first thing we notice is that Dh is the dominant motif at room temperature and remains as such up to temperatures close to melting. This shows that the global minimum structure is retained with at least 70\% occurrence up to ca. 450 K (see Table \ref{tab:fracPTMDprobHSA}). The next most abundant structures are twin, fcc with combined occurrence of $\sim$ 8\% at 300 K which increases to $\sim$ 18\% at 450 K. Between the two, the twin structures are marginally more abundant. We start observing icosahedral structures above 400 K. At 450 K, icosahedra are about 5\% which increases to 15\% at 500 K. The Au$_{147}$ clusters start melting close to 500 K and the amount of amorphous structures increases with further increase in the temperature. Before complete melting, a significant amount of \emph{mix} structures is observed (about 36\% at $\sim$ 500 K). These structures have features of more than a single motif and we will treat them in greater detail in Sec.~\ref{sec:mixed}. Finally, we remark that although the PTMD simulations are initialized with non-icosahedral structures, icosahedra are still observed indicating a good sampling of phase space.

We now make a comparison of PTMD and HSA. We use an energy cutoff of 1.5 eV above the global minimum when collecting the local minima from the PTMD simulations. Amorphous structures are not found within this energy range and hence are not considered while calculating the temperature-dependent probability of the various structures. The results are shown in Fig.~\ref{fgr:ptmdAu147} for a broader range of temperatures than for PTMD ($50<T<800$~K). At low temperatures ($<400$~K), we find that there is a good agreement between PTMD and HSA. In both cases, Dh is the dominant motif with little amount of fcc and twin (see Table~\ref{tab:fracPTMDprobHSA} for a quantitative comparison). Differences start appearing at higher temperatures ($>$ 400 K). HSA severely underestimates the amount of Ih and \emph{mix} structures at temperatures above 400 K. At 455 K and 499 K, 0.17\% and 1.33\% of Ih structures are expected from HSA calculations. However, the corresponding values observed from PTMD simulations are 5.18\% and 14.97\%. At 499 K, 36\% of \emph{mix} structures are observed in PTMD, while, only 0.20\% is expected based on HSA. These results clearly show that HSA cannot capture melting and is generally not reliable at  temperatures higher than ca.~$400$~K, most likely due to anharmonic effects.

Before analyzing the other cluster sizes, we discuss our results in relation to a recent work\cite{schebarchov2018AuHSA} which also employed HSA to study the finite-temperature structures of Au$_{147}$. The authors used Gupta potential with the same parameters used in the current work but they did not impose a distance cutoff on the interaction potential (in principle the cutoff should be imposed because it was used when fitting the potential parameters, but this introduces rather small differences since the interaction between Au atoms very quickly decreases with distance). We also note that the classification method is somewhat different from ours. Notwithstanding these differences, we find that our HSA results agree reasonably well with the calculations of \citet{schebarchov2018AuHSA}. This agreement combined with the fact that we collected the local minima from the PTMD simulations \emph{without} adding any minimum collected from global optimization runs, gives additional validation to the PTMD method as a means to effectively sample phase space.

\subsection{Au$_{201}$}
\begin{figure}[!t]
\centering
  \includegraphics[width=8.5cm]{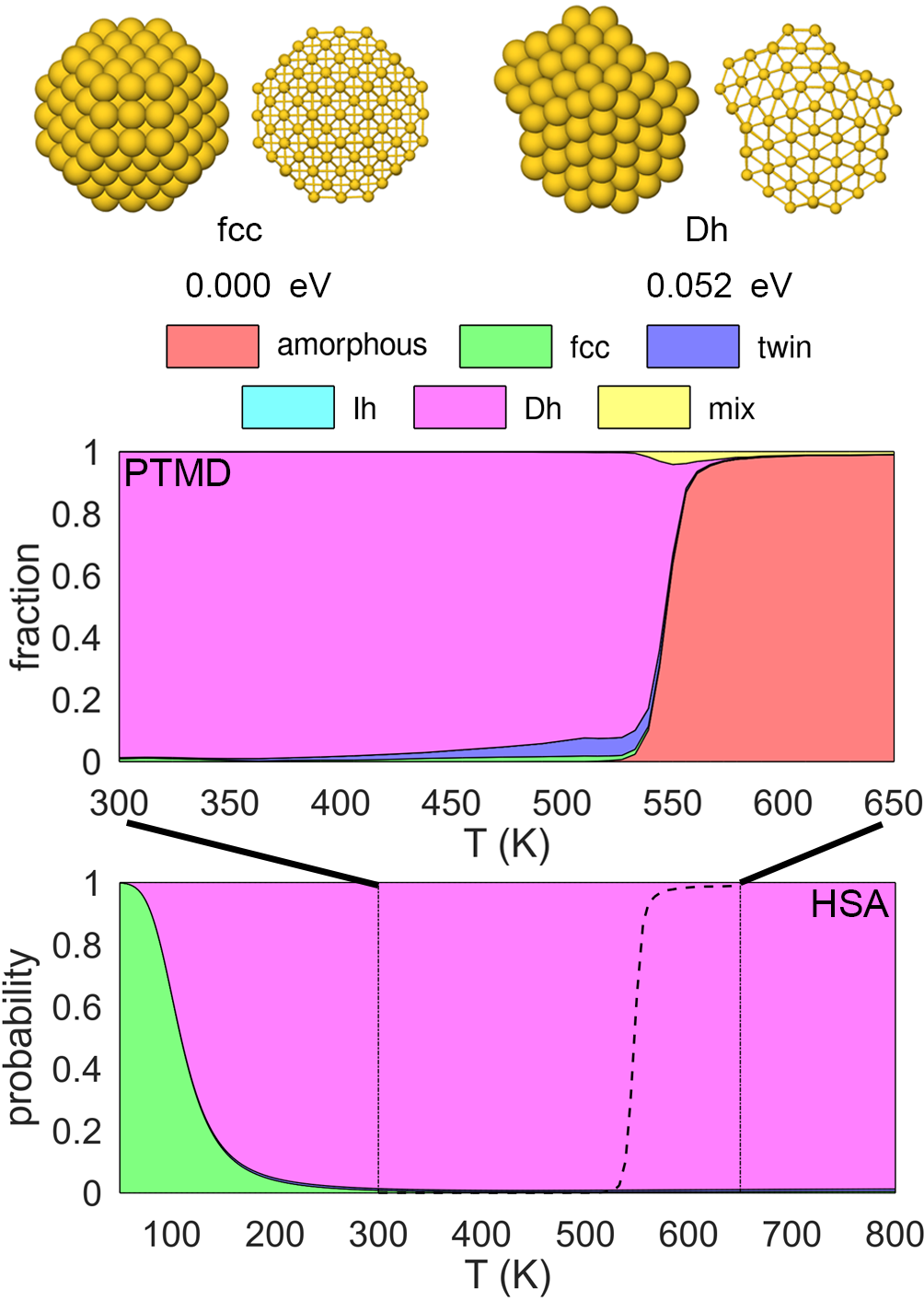}
  \caption{(top) The initial structures considered for PTMD simulations of Au$_{201}$ clusters along with their energy. The energy is measured relative to the global minimum. (middle) Occurrence fraction of various structures of Au$_{201}$ nanoclusters as a function of temperature in the range 300 K $-$ 650 K obtained via PTMD simulations. (bottom) Probability of various structures of Au$_{201}$ nanoclusters as a function of temperature in the range 50 K $-$ 800 K obtained via HSA. For comparison, we report with vertical lines  the range of PTMD temperatures and with a dashed line the fraction of amorphous structures calculated from PTMD simulations.}
  \label{fgr:ptmdAu201}
\end{figure}

The size 201 is a geometric ``magic'' size corresponding to a regular truncated octahedron which is also the global minimum (Fig.~\ref{fgr:ptmdAu201}). Bao \emph{et al.}, however, found a decahedron which is 7 meV lower in energy than the truncated octahedron although they use the same potential parameters that we have used. We believe that the discrepancy is possibly due to the difference in the distance cutoff used for the potential. In any case, the energy difference is very small and these structures can be treated as quasi-degenerate. We found a couple of decahedra which are higher in energy by 0.052 eV (shown in Fig. \ref{fgr:ptmdAu201}) and 0.056 eV. The truncated octahedron and the lowest energy decahedron were chosen as the initial structures for the PTMD simulations of Au$_{201}$ clusters. We verified that also in this case results become independent of the initial conditions after an initial phase of 0.5 $\mu$s. The PTMD production runs were carried out for an additional 2.0 $\mu$s.  We used an energy cutoff of 1.0 eV for collecting the local minima used in HSA.

Results of the PTMD simulations and HSA calculations for Au$_{201}$ clusters are shown in Fig. \ref{fgr:ptmdAu201}. From the PTMD simulations, we find that Dh is the dominant motif at room temperature and above. This is surprising since the global minimum is a truncated octahedron, but the reason will become clear after we consider the HSA results. There is a solid-solid transition from fcc (truncated octahedra) to decahedral structures well below  room temperature (between 100 K and 200 K) indicating that decahedra have high vibrational entropy. Due to this solid-solid transition, the probability of fcc and twin structures is almost zero at room temperature. Dh remains the most abundant motif up to melting. At temperatures greater than 400 K, small amounts of twin structures are observed in PTMD simulations. Very small amounts of \emph{mix} structures are observed close to melting (550 K). Unlike Au$_{147}$ clusters, icosahedral structures are completely absent at this size. Again, we find that there is very good agreement between PTMD and HSA up to ca. 500 K. However, HSA predicts almost zero twin/fcc structures whereas, non-negligible amount of twin/fcc structures are observed in PTMD simulations. 

The results of Au$_{201}$ show that relying solely on global minima to make predictions of the expected structures can be misleading and that it is important to evaluate the temperature-dependent structures, even  close to room temperature.
On the other hand, the low-temperature structural transition from fcc to Dh would be computationally very expensive to capture by means of PTMD alone -- in this case global optimization and HSA analysis are the most appropriate approaches. The latter are able to explore such an elusive low-energy part of the landscape with the additional advantage of including potentially relevant quantum corrections.

\begin{figure}[!b]
\centering
  \includegraphics[width=8.5cm]{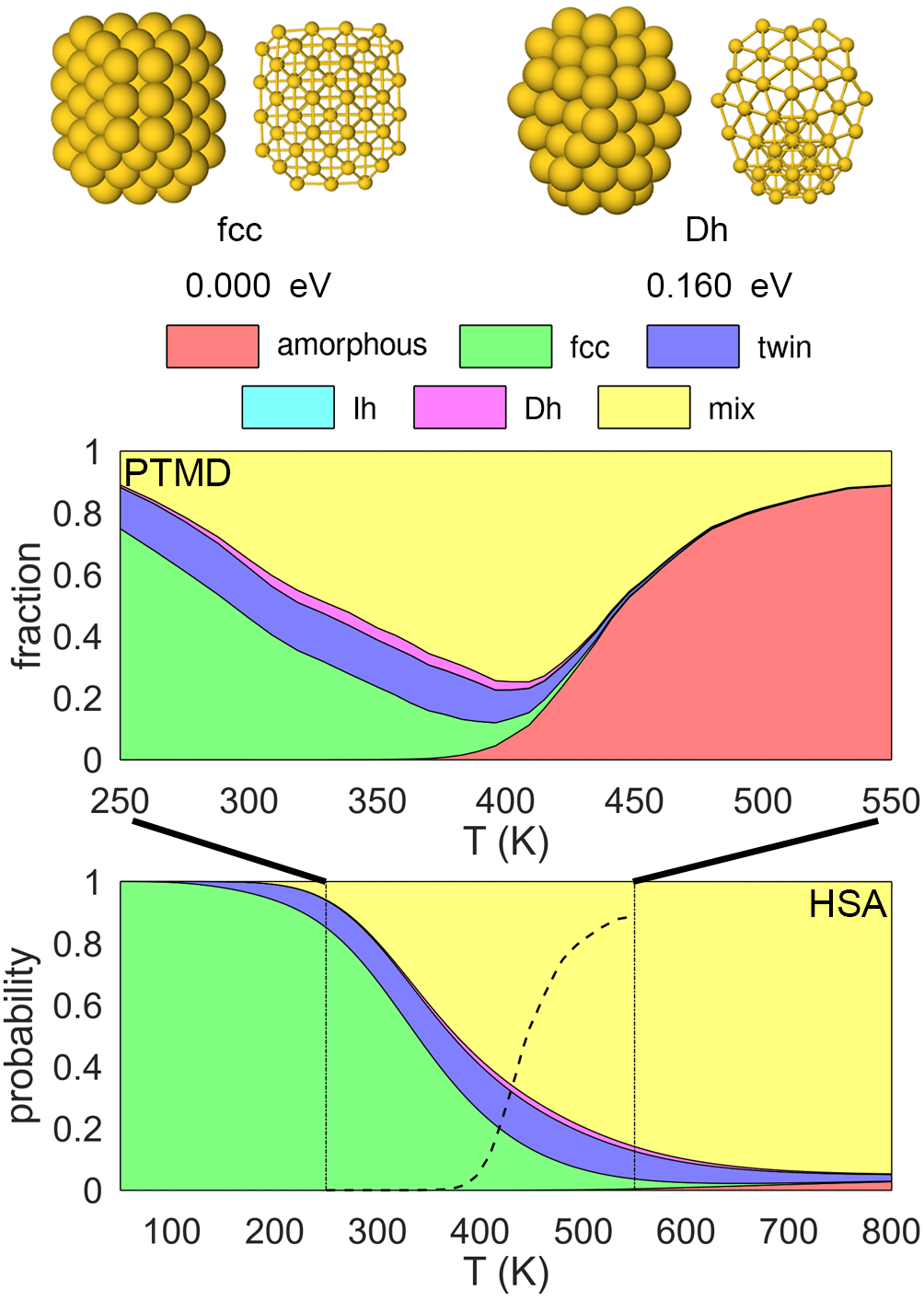}
  \caption{(top) The initial structures considered for PTMD simulations of Au$_{90}$ clusters along with their energy. The energy is measured relative to the global minimum. (middle) Occurrence fraction of various structures of Au$_{90}$ nanoclusters as a function of temperature in the range 250 K $-$ 550 K obtained via PTMD simulations. (bottom) Probability of various structures of Au$_{90}$ nanoclusters as a function of temperature in the range 50 K $-$ 800 K obtained via HSA. For comparison, we report with vertical lines  the range of PTMD temperatures and with a dashed line the fraction of amorphous structures calculated from PTMD simulations.}
  \label{fgr:ptmdAu90}
\end{figure}

\begin{figure*}[!t]
\centering
  \includegraphics[width=15cm]{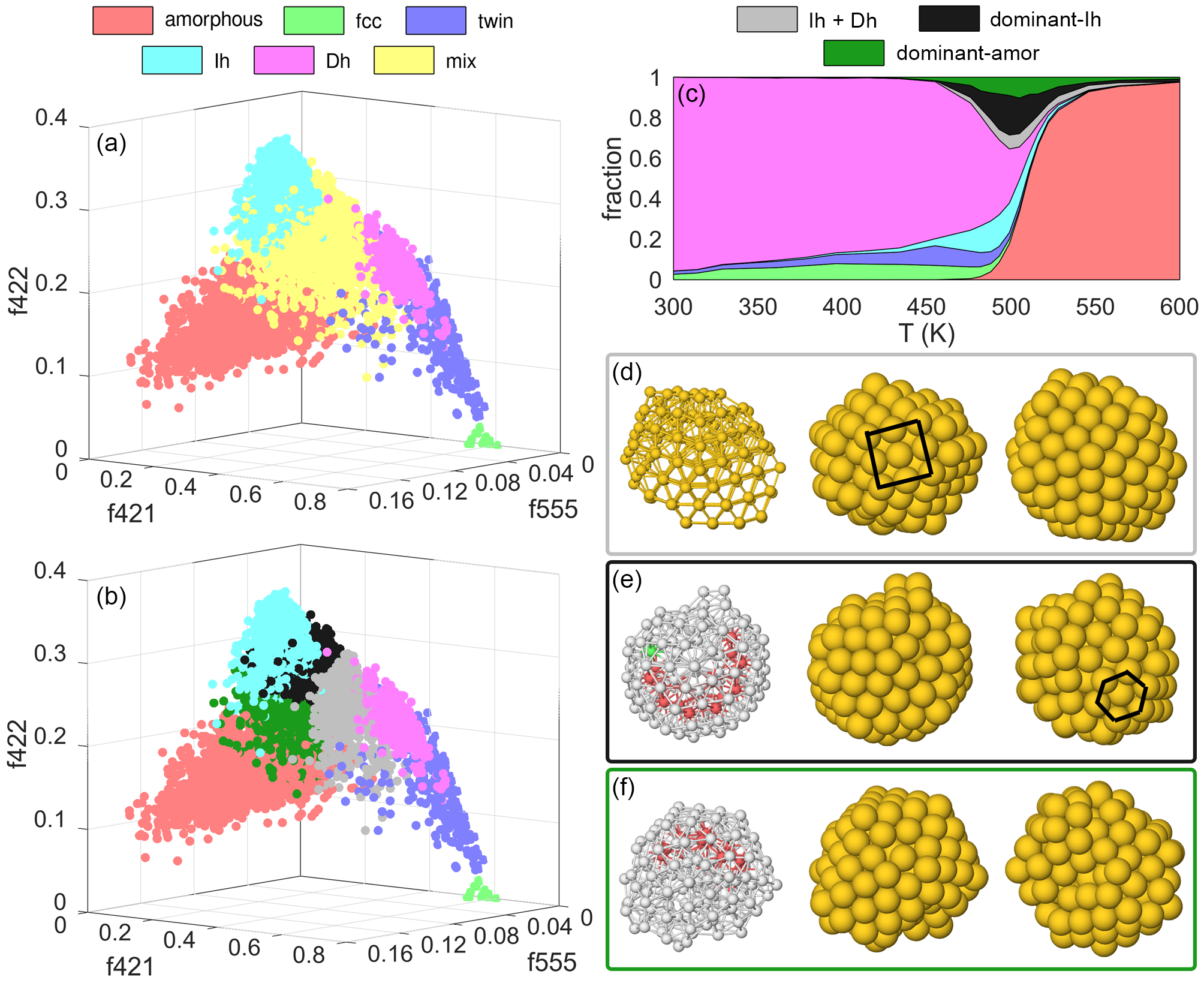}
  \caption{(a,b) 3D plot showing the CNA signatures of various structure types of Au$_{147}$ at 499 K. f421, f422, and f555 are the fractions of (421), (422), and (555) signatures, respectively. In (b), the structure family corresponding to the \emph{mix} structure is decomposed into three sub families. (c) Occurrence fraction of various structures of Au$_{147}$ nanoclusters as a function of temperature in the range 300 K $-$ 600 K obtained via PTMD simulations. The \emph{mix} structure type is decomposed into the three structure types described in the main text. Illustration of structures belonging to (d) \emph{Ih + Dh}, (e) \emph{dominant-Ih}, and (f) \emph{dominant-amor} structure types. In panel (d), the \{100\} facet is indicated by a square. In panel (e), the surface ``rosette'' defect is indicated by a hexagon. The atoms in the first figure in panels (e) and (f) are colored according to their coordination: pink indicates \emph{hcp} coordination, green \emph{fcc} coordination, white undefined coordination.} 
  \label{fgr:3DplotCNA147}
\end{figure*}

\begin{figure*}[!t]
\centering
  \includegraphics[width=15cm]{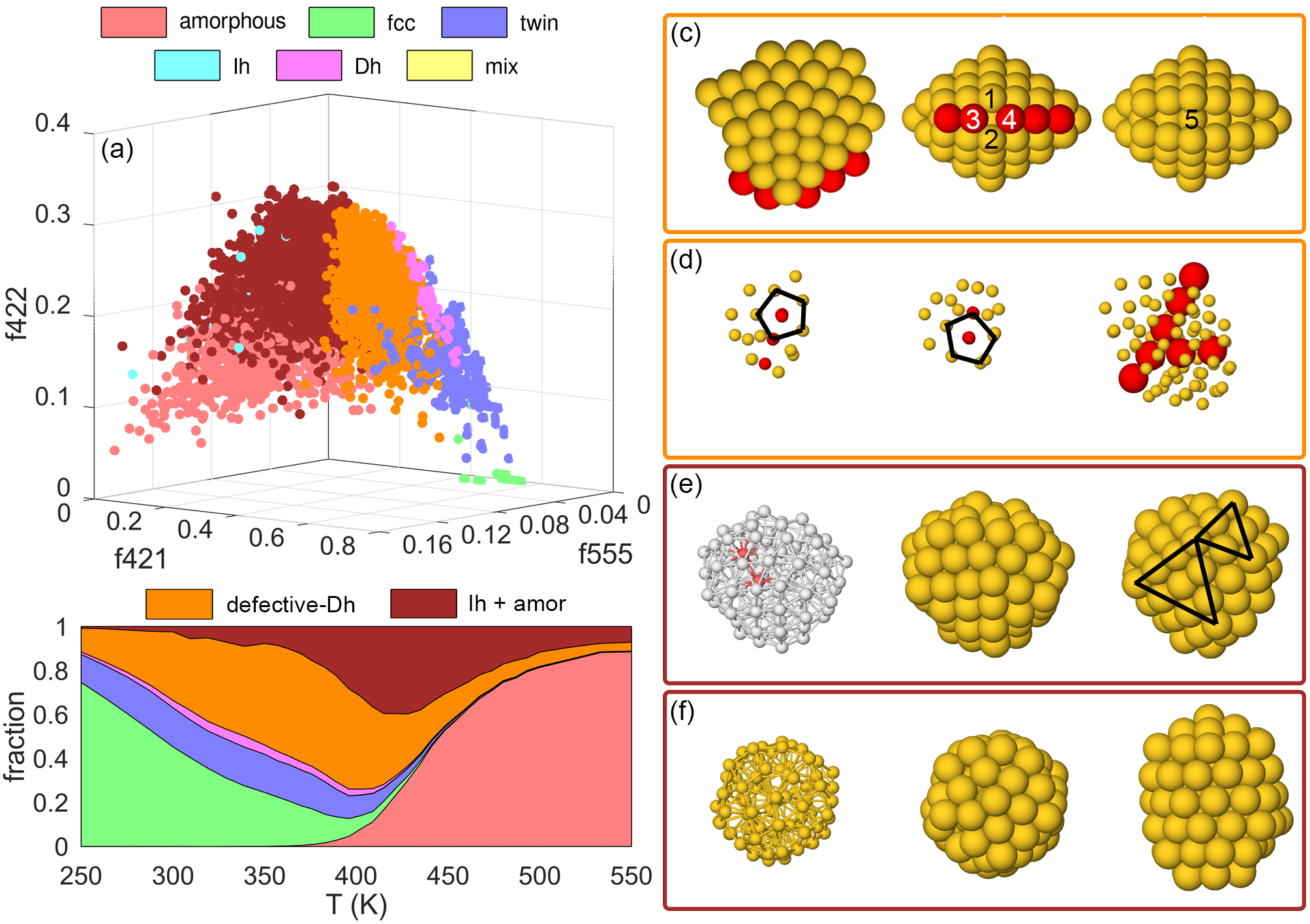}
  \caption{(a) 3D plot showing the CNA signatures of various structure types of Au$_{90}$ at 402 K. f421, f422, and f555 are the fractions of (421), (422), and (555) signatures, respectively. The structure family corresponding to the \emph{mix} structure is decomposed into two sub families. (b) Occurrence fraction of various structures of Au$_{90}$ nanoclusters as a function of temperature in the range 250 K $-$ 550 K obtained from PTMD simulations with Dh (+ hcp island) initial structure. The \emph{mix} structure type is decomposed into two structure types  described in the main text. Illustration of structures belonging to (c,d) \emph{defective-Dh} and (e,f) \emph{Ih + amor} structure types. (c) The rows of atoms indicated in red are pulled inside resulting in an increase of the distance between atoms 1 and 2. (d) Polydecahedron (poly-Dh) having two decahedral axes. Atoms belonging to the decahedral axes are colored in red. In the first two images the core is visualized along the two decahedral axes. (e) Structure having icosahedral and amorphous features. The triangles indicate {111}-like triangular facets with edges of length 3 and 4. (f) Structure resembling a 71-atom chiral structure. Pink indicates \emph{hcp} coordination, green \emph{fcc} coordination, and white undefined coordination.} 
  \label{fgr:3DplotCNA90}
\end{figure*}

\subsection{Au$_{90}$}
We now consider the size of 90 which is not a geometrical ``magic'' size. The global minimum is an fcc truncated octahedron structure. Close to this we find a twin which is 0.052 eV higher than the global minimum. The best decahedron is 0.143 eV higher and we also find a decahedron with an \emph{hcp} island which is 0.160 eV higher than the global minimum. In this case, we carried out the PTMD simulations for 2.6 $\mu$s with the fcc (global minimum) and the decahedron + \emph{hcp} island as the initial structures (Fig. \ref{fgr:ptmdAu90}). We used an energy cutoff of 1.0 eV for collecting the local minima. 

The PTMD and HSA results for Au$_{90}$ are shown in Fig.~\ref{fgr:ptmdAu90}. The global minimum motif, which is \emph{fcc}, remains dominant up to room temperature. As the temperature increases to $\sim$  310 K, the the most abundant structure changes from \emph{fcc} to \emph{mix}. In the range 300 K -- 380 K, we observe  $\sim$ 4\% Dh and  $\sim$ 15\% twin structures which remain almost constant in this temperature range. The clusters start melting around 400 K; at this temperature, the amount of \emph{mix} structures is maximum. In contrast to Au$_{147}$ and Au$_{201}$ clusters, the amount of \emph{mix} structures is higher in the case of Au$_{90}$ clusters. We look at these mixed structures more closely in the next section. The HSA results agree with the PTMD simulations up to ca. $\sim$ 400 K where the cluster does not melt yet.

\subsection{Mixed structures}
\label{sec:mixed}
In the experimental study of nanoclusters, a good portion of the structures is always difficult to classify. While calculating the fraction, such structures are classified as unidentified or ambiguous.\cite{wang2012AuImaging,young2010AuImagingInSitu,wells2015AuImagingACFraction,foster2018AuImagingACFraction} Given the strong structure-property relationship, it could prove to be useful to describe these structures in greater detail specially when they contribute significantly to the overall fraction. In the case of Au$_{90}$ and Au$_{147}$ clusters, we have already seen that \emph{mix} structures make a significant contribution. The amount of \emph{mix} structures is highest at size 90, accounting for up to ca. 60\% of the structures. It is lower at the size of 147 where it is mainly observed before melting and it is almost non-existent at 201. A similar size-dependent trend was reported in a previous HSA study\cite{schebarchov2018AuHSA} where the amount of ``ambiguous'' structures reduced as the size increased from 55 to 147.

We now focus on the \emph{mix} structures to study their make up. For this, we consider the fraction of (421), (422), and (555) CNA signatures denoted by f421, f422, and f555 respectively. In Fig.~\ref{fgr:3DplotCNA147}a, we construct a 3D plot with f421, f422, and f555 as the axes for the structures of Au$_{147}$ at 499 K obtained from PTMD simulations. Each structure type is represented by a different color on this scatter plot. The \emph{mix} structure family is surrounded by the Dh, Ih, and amorphous ones. A visual inspection of the structures belonging to the \emph{mix} family revealed that there were mainly three substructures: (i) structures having features of icosahedron and decahedron (Ih + Dh), (ii) structures which are predominantly icosahedron with some disordered ones (dominant-Ih), and (iii) structures which are predominantly amorphous with some icosahedral features (dominant-amor). Based on this preliminary classification, the \emph{mix} family was decomposed into three sub families using the k-means algorithm\cite{kmeans1979}$^,$\cite{kmeans1982}; results are reported in Fig.~\ref{fgr:3DplotCNA147}b.

An example of an \emph{Ih + Dh} structure is shown in Fig.~\ref{fgr:3DplotCNA147}d. The \{100\} facet is marked with a square (second image) which corresponds to a decahedron. At the same time, we can also observe the icosahedral features in the third image. The other two sub families (dominant-Ih, dominant-amor) are a combination of icosahedron and amorphous structures, the difference being the proportion of the icosahedral and amorphous features. In Fig.~\ref{fgr:3DplotCNA147}e, a dominant-Ih structure is shown. In the first image the atoms are colored according to their coordination. This image reveals that the majority of the structure has icosahedral features along with some amorphous region. In addition, these structures have ``rosette'' defects\cite{apra2004AuRosette} on the surface (marked with a hexagon in the third image). In contrast, the dominant-amor structures have higher fraction of amorphous features (see Fig. \ref{fgr:3DplotCNA147}f) and generally do not have surface ``rosette'' defects.

Interestingly, all the three sub families which make up the \emph{mix} family contain icosahedral features to a varying degree. 
In Fig.~\ref{fgr:3DplotCNA147}c, we re-plot the fraction of various structures as a function of temperature including these new structure types. The majority of the \emph{mix} family consists of the \emph{dominant-Ih} structure. As a result, the amount of icosahedral structures in the vicinity of the melting temperature is much higher than anticipated solely based on the icosahedral structures that have a \emph{central-Ih} coordinated atom (the structure family show in cyan).

Moving on to size 90, we observe that the icosahedral family is missing. Hence, we decompose the \emph{mix} structures into two sub families. In Fig.~\ref{fgr:3DplotCNA90}a, we construct a similar 3D plot for the Au$_{90}$ structures obtained from the PTMD simulations at 402 K: (i) defective decahedral structures (\emph{defective-Dh}, shown in orange) and (ii) structures having icosahedral and amorphous features (\emph{Ih + amor}, shown in brown). Referring to Fig. \ref{fgr:3DplotCNA90}b, at intermediate temperatures ($\sim$ 325 K to 400 K), \emph{defective-Dh} is the dominant structure type. Close to the melting, \emph{Ih + amor} becomes the dominant structure type. With increasing temperature, we observe the following sequence of dominant structures: fcc $\rightarrow$ defective-Dh $\rightarrow$ \emph{Ih + amor} $\rightarrow$ amorphous.

The decahedral structures at the size 90 can be formed by adding 15 atoms to the 75-atom Marks decahedron. There are mainly three kinds of defective-Dh structures. The most common among them is shown in Fig. \ref{fgr:3DplotCNA90}c. In this structure, the row of atoms that are on the \{100\} facets (marked in red) penetrate inwards. At the same time, the atoms 1 and 2 are pushed outwards and the distance between them increases. The gap left due to penetration of the outer rows of atoms is clearly visible in the third image. As a result, new bonds are formed between the atoms 3, 4 and the inner atom 5. Typically, such a defect appears only at the end of one of the twin planes. The second type of defective-Dh structure is the poly-decahedron\cite{polyDh2007} (shown in Fig.~\ref{fgr:3DplotCNA90}d) which has more than one decahedral axis. In the first two images, the core is visualized along the two decahedral axes (the pentagon indicates the decahedral axis). The two decahedral axes are clearly visible (marked in red) in the third image. The third defective-Dh structure has \emph{hcp} islands on the surface. An example of this structure is shown in the initial structure (on the right) in Fig.~\ref{fgr:ptmdAu90}. 

The \emph{Ih + amor} family also consists of a variety of icosahedral-based structures. The structure shown in Fig.~\ref{fgr:3DplotCNA90}e is a mixture of icosahedral and amorphous structures. The size 90 is intermediate between the icosahedral ``magic'' sizes 55 and 147. As a result, the triangular \{111\} facets corresponding to 55-atom and 147-atom icosahedra can co-exist on the surface (these are indicated by triangles in the third image of Fig.~\ref{fgr:3DplotCNA90}e). In Fig.~\ref{fgr:3DplotCNA90}f, an icosahedral structure is shown which is formed by adding 19 atoms to the 71-atom chiral structure.\cite{kostko2007chiral71}$^,$\cite{rapallo2012chiral71AuCo} In addition to these, we also observed structures resembling polyicosahedra.\cite{rossi2004pIh} 

In this section, we have used k-means algorithm to shed light on the \emph{mix} structures. The main objective is to illustrate the broad variety of structures that constitute \emph{mix} family. However, sophisticated methods (such as machine learning models) may be required for systems that exhibit more complex structures than those observed here.

\section{Conclusions}
In this paper, we determined the probabilities of the different structural motifs of Au$_{90}$, Au$_{147}$, and Au$_{201}$ nanoclusters, in the whole temperature range from 0 K to full melting. The numerical approach instrumental to obtain these results was based on parallel tempering molecular dynamics (PTMD) simulations, complemented by global optimization (GO) searches and Harmonic Superposition Approximation (HSA) calculations.

As a first result of general character, our simulations have shown that the equilibrium structures at finite temperature cannot be predicted on the basis of the global minimum alone, even below room temperature. Our GO searches have found that the global minimum of Au$_{147}$ is a a decahedral structure, whereas \emph{fcc} truncated octahedra are the global minima for Au$_{90}$ and Au$_{201}$ clusters. In fact, the finite-temperature results have shown that at size 147 decahedra dominate up to temperatures close to melting, but Au$_{201}$ undergoes a solid-solid transition \emph{fcc} $\rightarrow$ decahedron at $T< 200$~K. As a result, decahedron is the dominant motif at room temperatures for size 201 as well. In the case of Au$_{90}$, the global minimum \emph{fcc} remains dominant up to $\sim$ 325 K, and then other structures become more likely.

Along with the regular geometrical structures (fcc, twin, Ih, and Dh) we also found ``mixed'' or ``defective'' structures which we denote as the \emph{mix} structure family. These structures appear mainly in Au$_{90}$ and Au$_{147}$ nanoclusters. At size 147, all the \emph{mix} structures contain partial icosahedral features. Based on the fraction of (421), (422), and (555) CNA signatures, the \emph{mix} family was classified into three sub families having different proportions of icosahedral features. Considering the \emph{mix} structures, icosahedron becomes the dominant structure just before melting (490 K $-$ 500 K). The dominant structures at size 147 thus follow the  sequence: Dh $\rightarrow$ Ih (including \emph{mix}) $\rightarrow$ amorphous. Although Au nanoclusters energetically disfavor the icosahedral motif, these structures are observed at higher temperatures. Icosahedra start appearing at temperatures greater than 400 K and combined with \emph{mix} structures; their fraction continually increases up to 500 K where they are the most abundant structure. The fraction of \emph{mix} structures is significantly higher for Au$_{90}$. In this case, the \emph{mix} structural family was decomposed into two sub families. One of them is made of defective decahedra. The remaining \emph{mix} structures are a combination of icosahedral and amorphous structures. In the case of Au$_{90}$ clusters, we observe the following structural changes: \emph{fcc} (up to $\sim$ 320 K) $\rightarrow$ defective decahedra (320 K $-$ 410 K) $\rightarrow$ Ih + amorphous (410 K $-$ 435 K) $\rightarrow$ amorphous.

In all cases, we have found wide temperature ranges (between 300 and 400 K at least) in which there is a very good agreement between the results of PTMD simulations and of HSA calculations. Beyond 400 K, we start observing increasing differences between PTMD and HSA. In the case of Au$_{147}$, HSA does not predict Ih structures before the onset of melting. However, Ih structures start appearing already at 400 K in the PTMD simulations. Hence, PTMD is accurate from room temperature up to melting. On the other hand, HSA is suitable to study the low temperature (less than room temperature) solid-solid structural transitions, such as the \emph{fcc} $\rightarrow$ Dh transition found for Au$_{201}$ which occurs at $T< 200$~K. While it is difficult to capture the low temperature solid-solid transitions using PTMD directly, the HSA method can be applied to the local minima collected from the PTMD simulations (enriched by GO searches if necessary). In our cases, the low temperature solid-solid transitions were studied by using the local minima from PTMD alone without the need of additional minima from optimization searches, since the basin hopping searches confirmed the low-energy structures found by PTMD. As a final comment, we confirm that the proposed method combining PTMD and HSA is capable of an accurate description of the probabilities of the different motifs in the whole temperature range for systems in which the competition between the motifs is far from trivial.

Given the simplicity and the robustness of our method, we believe that it would be transferable to larger clusters and to nanoalloys. This will be taken up as future research. The method can be used in principle with any type of force field, and also with \textit{ab initio} calculations. Very often, in the process of identifying and designing optimal structures for a given property (such as catalytic activity), idealized geometrical shapes are considered for screening a large set of structures. Also, the proportion of various geometrical motifs are usually not considered nor its changes with the operation temperature. We believe that using the information of structural distribution as a function of temperature will greatly improve the screening process.

\section*{Conflicts of interest}
There are no conflicts to declare.

\section*{Acknowledgements}
This work has been supported by the project ``Understanding and Tuning FRiction through nanOstructure Manipulation (UTFROM)'' funded by MIUR  Progetti di Ricerca di Rilevante Interesse Nazionale (PRIN) Bando 2017 - grant 20178PZCB5. R.F. acknowledges the Progetto di Eccellenza of the Physics Department of the University of Genoa for financial and the International Research Network Nanoalloys of CNRS for networking support. The authors acknowledge PRACE for awarding us access to Marconi100 at CINECA, Italy.





\bibliography{rsc} 

\providecommand*{\mcitethebibliography}{\thebibliography}
\csname @ifundefined\endcsname{endmcitethebibliography}
{\let\endmcitethebibliography\endthebibliography}{}
\begin{mcitethebibliography}{85}
\providecommand*{\natexlab}[1]{#1}
\providecommand*{\mciteSetBstSublistMode}[1]{}
\providecommand*{\mciteSetBstMaxWidthForm}[2]{}
\providecommand*{\mciteBstWouldAddEndPuncttrue}
  {\def\EndOfBibitem{\unskip.}}
\providecommand*{\mciteBstWouldAddEndPunctfalse}
  {\let\EndOfBibitem\relax}
\providecommand*{\mciteSetBstMidEndSepPunct}[3]{}
\providecommand*{\mciteSetBstSublistLabelBeginEnd}[3]{}
\providecommand*{\EndOfBibitem}{}
\mciteSetBstSublistMode{f}
\mciteSetBstMaxWidthForm{subitem}
{(\emph{\alph{mcitesubitemcount}})}
\mciteSetBstSublistLabelBeginEnd{\mcitemaxwidthsubitemform\space}
{\relax}{\relax}

\bibitem[Cunningham \emph{et~al.}(1998)Cunningham, Vogel, Kageyama, Tsubota,
  and Haruta]{haruta1998AuCat}
D.~H. Cunningham, G.~W. Vogel, H.~Kageyama, S.~Tsubota and M.~Haruta, \emph{J.
  Catal.}, 1998, \textbf{177}, 1--10\relax
\mciteBstWouldAddEndPuncttrue
\mciteSetBstMidEndSepPunct{\mcitedefaultmidpunct}
{\mcitedefaultendpunct}{\mcitedefaultseppunct}\relax
\EndOfBibitem
\bibitem[Wallace and Whetten(2000)]{whetten2000AuCat}
W.~T. Wallace and R.~L. Whetten, \emph{J. Phys. Chem. B}, 2000, \textbf{104},
  10964--10968\relax
\mciteBstWouldAddEndPuncttrue
\mciteSetBstMidEndSepPunct{\mcitedefaultmidpunct}
{\mcitedefaultendpunct}{\mcitedefaultseppunct}\relax
\EndOfBibitem
\bibitem[Molina and Hammer(2005)]{rev2005AuCat}
L.~M. Molina and B.~Hammer, \emph{Appl. Catal. A: Gen.}, 2005, \textbf{291},
  21--31\relax
\mciteBstWouldAddEndPuncttrue
\mciteSetBstMidEndSepPunct{\mcitedefaultmidpunct}
{\mcitedefaultendpunct}{\mcitedefaultseppunct}\relax
\EndOfBibitem
\bibitem[Hashmi and Hutchings(2006)]{rev2006AuCat}
A.~S.~K. Hashmi and G.~J. Hutchings, \emph{Angew. Chem. Int. Ed.}, 2006,
  \textbf{45}, 7896--7936\relax
\mciteBstWouldAddEndPuncttrue
\mciteSetBstMidEndSepPunct{\mcitedefaultmidpunct}
{\mcitedefaultendpunct}{\mcitedefaultseppunct}\relax
\EndOfBibitem
\bibitem[Chen and Chen(2009)]{chen2009AuCat}
W.~Chen and S.~Chen, \emph{Angew. Chem. Int. Ed.}, 2009, \textbf{48},
  4386--4389\relax
\mciteBstWouldAddEndPuncttrue
\mciteSetBstMidEndSepPunct{\mcitedefaultmidpunct}
{\mcitedefaultendpunct}{\mcitedefaultseppunct}\relax
\EndOfBibitem
\bibitem[Yamamoto \emph{et~al.}(2012)Yamamoto, Yano, Kouchi, Obora, Arakawa,
  and Kawasaki]{kawasaki2012AuCat}
H.~Yamamoto, H.~Yano, H.~Kouchi, Y.~Obora, R.~Arakawa and H.~Kawasaki,
  \emph{Nanoscale}, 2012, \textbf{4}, 4148--4154\relax
\mciteBstWouldAddEndPuncttrue
\mciteSetBstMidEndSepPunct{\mcitedefaultmidpunct}
{\mcitedefaultendpunct}{\mcitedefaultseppunct}\relax
\EndOfBibitem
\bibitem[Zhao and Jin(2018)]{rev2018AuCatAPCs}
J.~Zhao and R.~Jin, \emph{Nano Today}, 2018, \textbf{18}, 86--102\relax
\mciteBstWouldAddEndPuncttrue
\mciteSetBstMidEndSepPunct{\mcitedefaultmidpunct}
{\mcitedefaultendpunct}{\mcitedefaultseppunct}\relax
\EndOfBibitem
\bibitem[Turner \emph{et~al.}(2008)Turner, Golovko, Vaughan, Abdulkin,
  Berenguer-Murcia, Tikhov, Johnson, and Lambert]{turner2008AuCatSizeEffect}
M.~Turner, V.~B. Golovko, O.~P.~H. Vaughan, P.~Abdulkin, A.~Berenguer-Murcia,
  M.~S. Tikhov, B.~F.~G. Johnson and R.~M. Lambert, \emph{Nature}, 2008,
  \textbf{454}, 981--983\relax
\mciteBstWouldAddEndPuncttrue
\mciteSetBstMidEndSepPunct{\mcitedefaultmidpunct}
{\mcitedefaultendpunct}{\mcitedefaultseppunct}\relax
\EndOfBibitem
\bibitem[Li \emph{et~al.}(2015)Li, Li, Pedersen, Gao, Khetrapal, Jonsson, and
  Zeng]{li2015AuCatShapeEffect}
H.~Li, L.~Li, A.~Pedersen, Y.~Gao, N.~Khetrapal, H.~Jonsson and X.~C. Zeng,
  \emph{Nano Lett.}, 2015, \textbf{15}, 682--688\relax
\mciteBstWouldAddEndPuncttrue
\mciteSetBstMidEndSepPunct{\mcitedefaultmidpunct}
{\mcitedefaultendpunct}{\mcitedefaultseppunct}\relax
\EndOfBibitem
\bibitem[Daniel and Astruc(2004)]{rev2004AuBioSens}
M.~Daniel and D.~Astruc, \emph{Chem. Rev.}, 2004, \textbf{104}, 293--346\relax
\mciteBstWouldAddEndPuncttrue
\mciteSetBstMidEndSepPunct{\mcitedefaultmidpunct}
{\mcitedefaultendpunct}{\mcitedefaultseppunct}\relax
\EndOfBibitem
\bibitem[Eustis and El-Sayed(2005)]{rev2005AuSPR}
S.~Eustis and M.~A. El-Sayed, \emph{Chem. Soc. Rev.}, 2005, \textbf{35},
  209--217\relax
\mciteBstWouldAddEndPuncttrue
\mciteSetBstMidEndSepPunct{\mcitedefaultmidpunct}
{\mcitedefaultendpunct}{\mcitedefaultseppunct}\relax
\EndOfBibitem
\bibitem[Amendola \emph{et~al.}(2017)Amendola, Pilot, Frasconi, Marago, and
  Iati]{rev2017AuSPR}
V.~Amendola, R.~Pilot, M.~Frasconi, O.~M. Marago and M.~A. Iati, \emph{J.
  Phys.: Condens. Matter}, 2017, \textbf{29}, 203002\relax
\mciteBstWouldAddEndPuncttrue
\mciteSetBstMidEndSepPunct{\mcitedefaultmidpunct}
{\mcitedefaultendpunct}{\mcitedefaultseppunct}\relax
\EndOfBibitem
\bibitem[Loynachan \emph{et~al.}(2019)Loynachan, Soleimany, Dudani, Lin, Najer,
  Bekdemir, Chen, Bhatia, and Stevens]{colleen2019AuinVivoDisease}
C.~N. Loynachan, A.~P. Soleimany, J.~S. Dudani, Y.~Lin, A.~Najer, A.~Bekdemir,
  Q.~Chen, S.~N. Bhatia and M.~M. Stevens, \emph{Nat. Nanotechnol}, 2019,
  \textbf{14}, 883--890\relax
\mciteBstWouldAddEndPuncttrue
\mciteSetBstMidEndSepPunct{\mcitedefaultmidpunct}
{\mcitedefaultendpunct}{\mcitedefaultseppunct}\relax
\EndOfBibitem
\bibitem[Porret \emph{et~al.}(2020)Porret, Guevel, and
  Coll]{rev2020AuBioMedical}
E.~Porret, X.~L. Guevel and J.~Coll, \emph{J. Mater. Chem. B}, 2020,
  \textbf{8}, 2216--2232\relax
\mciteBstWouldAddEndPuncttrue
\mciteSetBstMidEndSepPunct{\mcitedefaultmidpunct}
{\mcitedefaultendpunct}{\mcitedefaultseppunct}\relax
\EndOfBibitem
\bibitem[Pyykko(2004)]{pyykko2004AuRelEffects}
P.~Pyykko, \emph{Angew. Chem. Int. Ed.}, 2004, \textbf{43}, 4412--4456\relax
\mciteBstWouldAddEndPuncttrue
\mciteSetBstMidEndSepPunct{\mcitedefaultmidpunct}
{\mcitedefaultendpunct}{\mcitedefaultseppunct}\relax
\EndOfBibitem
\bibitem[Garzon \emph{et~al.}(1998)Garzon, Michaelian, Beltran,
  Posada-Amarillas, Ordejon, Artacho, Sanchez-Portal, and
  Soler]{garzon1998AuRelSRAmor}
I.~L. Garzon, K.~Michaelian, M.~R. Beltran, A.~Posada-Amarillas, P.~Ordejon,
  E.~Artacho, D.~Sanchez-Portal and J.~M. Soler, \emph{Phys. Rev. Lett.}, 1998,
  \textbf{81}, 1600--1603\relax
\mciteBstWouldAddEndPuncttrue
\mciteSetBstMidEndSepPunct{\mcitedefaultmidpunct}
{\mcitedefaultendpunct}{\mcitedefaultseppunct}\relax
\EndOfBibitem
\bibitem[Apra \emph{et~al.}(2006)Apra, Ferrando, and
  Fortunelli]{apra2006AuRelSRAmor}
E.~Apra, R.~Ferrando and A.~Fortunelli, \emph{Phys. Rev. B}, 2006, \textbf{73},
  205414\relax
\mciteBstWouldAddEndPuncttrue
\mciteSetBstMidEndSepPunct{\mcitedefaultmidpunct}
{\mcitedefaultendpunct}{\mcitedefaultseppunct}\relax
\EndOfBibitem
\bibitem[Hakkinen and Landman(2000)]{hakkinen2000AuPlanarStructs}
H.~Hakkinen and U.~Landman, \emph{Phys. Rev. B}, 2000, \textbf{62},
  R2287--R2290\relax
\mciteBstWouldAddEndPuncttrue
\mciteSetBstMidEndSepPunct{\mcitedefaultmidpunct}
{\mcitedefaultendpunct}{\mcitedefaultseppunct}\relax
\EndOfBibitem
\bibitem[Bravo-Perez \emph{et~al.}(1999)Bravo-Perez, Garzon, and
  Novaro]{perez1999AuPlanar}
G.~Bravo-Perez, I.~L. Garzon and O.~Novaro, \emph{J. Mol. Struct. Theocem},
  1999, \textbf{493}, 225--231\relax
\mciteBstWouldAddEndPuncttrue
\mciteSetBstMidEndSepPunct{\mcitedefaultmidpunct}
{\mcitedefaultendpunct}{\mcitedefaultseppunct}\relax
\EndOfBibitem
\bibitem[Hakkinen \emph{et~al.}(2003)Hakkinen, Yoon, Landman, Li, Zhai, and
  Wang]{hakkinen2003AuPlanarStructs}
H.~Hakkinen, B.~Yoon, U.~Landman, X.~Li, H.~Zhai and L.~Wang, \emph{J. Phys.
  Chem. A}, 2003, \textbf{107}, 6168--6175\relax
\mciteBstWouldAddEndPuncttrue
\mciteSetBstMidEndSepPunct{\mcitedefaultmidpunct}
{\mcitedefaultendpunct}{\mcitedefaultseppunct}\relax
\EndOfBibitem
\bibitem[Hakkinen \emph{et~al.}(2002)Hakkinen, Moseler, and
  Landman]{hakkinen2002Au2Dto3DCuAg}
H.~Hakkinen, M.~Moseler and U.~Landman, \emph{Phys. Rev. Lett.}, 2002,
  \textbf{89}, 033401\relax
\mciteBstWouldAddEndPuncttrue
\mciteSetBstMidEndSepPunct{\mcitedefaultmidpunct}
{\mcitedefaultendpunct}{\mcitedefaultseppunct}\relax
\EndOfBibitem
\bibitem[Furche \emph{et~al.}(2002)Furche, Ahlrichs, Weis, Jacob, Gilb,
  Bierweiler, and Kappes]{furche2002AuPlanarStructsExpIon}
F.~Furche, R.~Ahlrichs, P.~Weis, C.~Jacob, S.~Gilb, T.~Bierweiler and M.~M.
  Kappes, \emph{J. Chem. Phys.}, 2002, \textbf{117}, 6982--6990\relax
\mciteBstWouldAddEndPuncttrue
\mciteSetBstMidEndSepPunct{\mcitedefaultmidpunct}
{\mcitedefaultendpunct}{\mcitedefaultseppunct}\relax
\EndOfBibitem
\bibitem[Johansson \emph{et~al.}(2004)Johansson, Sundholm, and
  Vaara]{johansson2004AuCageStructs}
M.~P. Johansson, D.~Sundholm and J.~Vaara, \emph{Angew. Chem. Int. Ed.}, 2004,
  \textbf{43}, 2678--2681\relax
\mciteBstWouldAddEndPuncttrue
\mciteSetBstMidEndSepPunct{\mcitedefaultmidpunct}
{\mcitedefaultendpunct}{\mcitedefaultseppunct}\relax
\EndOfBibitem
\bibitem[Gu \emph{et~al.}(2004)Gu, Ji, Wei, and Gong]{gu2004AuCageStructs}
X.~Gu, M.~Ji, S.~H. Wei and X.~G. Gong, \emph{Phys. Rev. B}, 2004, \textbf{70},
  205401\relax
\mciteBstWouldAddEndPuncttrue
\mciteSetBstMidEndSepPunct{\mcitedefaultmidpunct}
{\mcitedefaultendpunct}{\mcitedefaultseppunct}\relax
\EndOfBibitem
\bibitem[Fa and Dong(2006)]{fa2006AuCageStructs}
W.~Fa and J.~Dong, \emph{J. Chem. Phys.}, 2006, \textbf{124}, 114310\relax
\mciteBstWouldAddEndPuncttrue
\mciteSetBstMidEndSepPunct{\mcitedefaultmidpunct}
{\mcitedefaultendpunct}{\mcitedefaultseppunct}\relax
\EndOfBibitem
\bibitem[Xing \emph{et~al.}(2006)Xing, Yoon, Landman, and
  Parks]{xing2006AuCageStructs}
X.~Xing, B.~Yoon, U.~Landman and J.~H. Parks, \emph{Phys. Rev. B}, 2006,
  \textbf{74}, 165423\relax
\mciteBstWouldAddEndPuncttrue
\mciteSetBstMidEndSepPunct{\mcitedefaultmidpunct}
{\mcitedefaultendpunct}{\mcitedefaultseppunct}\relax
\EndOfBibitem
\bibitem[Ferrando \emph{et~al.}(2009)Ferrando, Barcaro, and
  Fortunelli]{riccardo2009AuFlatMgO}
R.~Ferrando, G.~Barcaro and A.~Fortunelli, \emph{Phys. Rev. Lett.}, 2009,
  \textbf{102}, 216102\relax
\mciteBstWouldAddEndPuncttrue
\mciteSetBstMidEndSepPunct{\mcitedefaultmidpunct}
{\mcitedefaultendpunct}{\mcitedefaultseppunct}\relax
\EndOfBibitem
\bibitem[Ferrando \emph{et~al.}(2011)Ferrando, Barcaro, and
  Fortunelli]{riccardo2011AuFlatMgO}
R.~Ferrando, G.~Barcaro and A.~Fortunelli, \emph{Phys. Rev. B}, 2011,
  \textbf{83}, 045418\relax
\mciteBstWouldAddEndPuncttrue
\mciteSetBstMidEndSepPunct{\mcitedefaultmidpunct}
{\mcitedefaultendpunct}{\mcitedefaultseppunct}\relax
\EndOfBibitem
\bibitem[Damianos and Ferrando(2012)]{damianos2012AuSteppedMgO}
K.~Damianos and R.~Ferrando, \emph{Nanoscale}, 2012, \textbf{4},
  1101--1108\relax
\mciteBstWouldAddEndPuncttrue
\mciteSetBstMidEndSepPunct{\mcitedefaultmidpunct}
{\mcitedefaultendpunct}{\mcitedefaultseppunct}\relax
\EndOfBibitem
\bibitem[Michaelian \emph{et~al.}(1999)Michaelian, Rendon, and
  Garzon]{michaelian1999AuRelSRAmor}
K.~Michaelian, N.~Rendon and I.~L. Garzon, \emph{Phys. Rev. B}, 1999,
  \textbf{60}, 2000--2010\relax
\mciteBstWouldAddEndPuncttrue
\mciteSetBstMidEndSepPunct{\mcitedefaultmidpunct}
{\mcitedefaultendpunct}{\mcitedefaultseppunct}\relax
\EndOfBibitem
\bibitem[Gupta(1981)]{gupta1981}
R.~P. Gupta, \emph{Phys. Rev. B}, 1981, \textbf{23}, 62--65\relax
\mciteBstWouldAddEndPuncttrue
\mciteSetBstMidEndSepPunct{\mcitedefaultmidpunct}
{\mcitedefaultendpunct}{\mcitedefaultseppunct}\relax
\EndOfBibitem
\bibitem[Rosato \emph{et~al.}(1989)Rosato, Guillope, and Legrand]{rgl1989}
V.~Rosato, M.~Guillope and B.~Legrand, \emph{Philos. Mag. A}, 1989,
  \textbf{59}, 321--336\relax
\mciteBstWouldAddEndPuncttrue
\mciteSetBstMidEndSepPunct{\mcitedefaultmidpunct}
{\mcitedefaultendpunct}{\mcitedefaultseppunct}\relax
\EndOfBibitem
\bibitem[Huang \emph{et~al.}(2008)Huang, ji, Dong, Gu, Wang, Gong, and
  Wang]{huang2008AuExpPESDFTnonIco}
W.~Huang, M.~ji, C.~Dong, X.~Gu, L.~Wang, X.~G. Gong and L.~Wang, \emph{ACS
  Nano}, 2008, \textbf{2}, 897--904\relax
\mciteBstWouldAddEndPuncttrue
\mciteSetBstMidEndSepPunct{\mcitedefaultmidpunct}
{\mcitedefaultendpunct}{\mcitedefaultseppunct}\relax
\EndOfBibitem
\bibitem[Bao \emph{et~al.}(2009)Bao, Goedecker, Koga, Lançon, and
  Neelov]{bao2009AuBHMC}
K.~Bao, S.~Goedecker, K.~Koga, F.~Lançon and A.~Neelov, \emph{Phys. Rev. B},
  2009, \textbf{79}, 041405(R)\relax
\mciteBstWouldAddEndPuncttrue
\mciteSetBstMidEndSepPunct{\mcitedefaultmidpunct}
{\mcitedefaultendpunct}{\mcitedefaultseppunct}\relax
\EndOfBibitem
\bibitem[Schebarchov \emph{et~al.}(2018)Schebarchov, Baletto, and
  Wales]{schebarchov2018AuHSA}
D.~Schebarchov, F.~Baletto and D.~J. Wales, \emph{Nanoscale}, 2018,
  \textbf{10}, 2004--2016\relax
\mciteBstWouldAddEndPuncttrue
\mciteSetBstMidEndSepPunct{\mcitedefaultmidpunct}
{\mcitedefaultendpunct}{\mcitedefaultseppunct}\relax
\EndOfBibitem
\bibitem[Nelli \emph{et~al.}(2020)Nelli, Rossi, Wang, Palmer, and
  Ferrando]{nelli2020AuBHMC}
D.~Nelli, G.~Rossi, Z.~Wang, R.~E. Palmer and R.~Ferrando, \emph{Nanoscale},
  2020, \textbf{12}, 7688--7699\relax
\mciteBstWouldAddEndPuncttrue
\mciteSetBstMidEndSepPunct{\mcitedefaultmidpunct}
{\mcitedefaultendpunct}{\mcitedefaultseppunct}\relax
\EndOfBibitem
\bibitem[Palomares-Baez \emph{et~al.}(2017)Palomares-Baez, Panizon, and
  Ferrando]{palomaresBaez2017AuDFT}
J.~Palomares-Baez, E.~Panizon and R.~Ferrando, \emph{Nano Lett.}, 2017,
  \textbf{17}, 5394--5401\relax
\mciteBstWouldAddEndPuncttrue
\mciteSetBstMidEndSepPunct{\mcitedefaultmidpunct}
{\mcitedefaultendpunct}{\mcitedefaultseppunct}\relax
\EndOfBibitem
\bibitem[Wang and Palmer(2012)]{wang2012AuImaging}
Z.~W. Wang and R.~E. Palmer, \emph{Phys. Rev. Lett.}, 2012, \textbf{108},
  245502\relax
\mciteBstWouldAddEndPuncttrue
\mciteSetBstMidEndSepPunct{\mcitedefaultmidpunct}
{\mcitedefaultendpunct}{\mcitedefaultseppunct}\relax
\EndOfBibitem
\bibitem[Young \emph{et~al.}(2010)Young, van Huis, Zandbergen, Xu, and
  Kirkland]{young2010AuImagingInSitu}
N.~P. Young, M.~A. van Huis, H.~W. Zandbergen, H.~Xu and A.~I. Kirkland,
  \emph{Ultramicroscopy}, 2010, \textbf{110}, 506--516\relax
\mciteBstWouldAddEndPuncttrue
\mciteSetBstMidEndSepPunct{\mcitedefaultmidpunct}
{\mcitedefaultendpunct}{\mcitedefaultseppunct}\relax
\EndOfBibitem
\bibitem[Wells \emph{et~al.}(2015)Wells, Rossi, Ferrando, and
  Palmer]{wells2015AuImagingACFraction}
D.~M. Wells, G.~Rossi, R.~Ferrando and R.~E. Palmer, \emph{Nanoscale}, 2015,
  \textbf{7}, 6498--6503\relax
\mciteBstWouldAddEndPuncttrue
\mciteSetBstMidEndSepPunct{\mcitedefaultmidpunct}
{\mcitedefaultendpunct}{\mcitedefaultseppunct}\relax
\EndOfBibitem
\bibitem[Foster \emph{et~al.}(2018)Foster, Ferrando, and
  Palmer]{foster2018AuImagingACFraction}
D.~M. Foster, R.~Ferrando and R.~E. Palmer, \emph{Nat. Commun.}, 2018,
  \textbf{9}, 1323\relax
\mciteBstWouldAddEndPuncttrue
\mciteSetBstMidEndSepPunct{\mcitedefaultmidpunct}
{\mcitedefaultendpunct}{\mcitedefaultseppunct}\relax
\EndOfBibitem
\bibitem[Franke \emph{et~al.}(1993)Franke, Hilf, and Borrmann]{hsa1993}
G.~Franke, E.~R. Hilf and P.~Borrmann, \emph{J. Chern. Phys.}, 1993,
  \textbf{98}, 3496--3502\relax
\mciteBstWouldAddEndPuncttrue
\mciteSetBstMidEndSepPunct{\mcitedefaultmidpunct}
{\mcitedefaultendpunct}{\mcitedefaultseppunct}\relax
\EndOfBibitem
\bibitem[Doye and Calvo(2001)]{doye2001hsaLJ}
J.~P.~K. Doye and F.~Calvo, \emph{Phys. Rev. Lett.}, 2001, \textbf{86},
  3570--3573\relax
\mciteBstWouldAddEndPuncttrue
\mciteSetBstMidEndSepPunct{\mcitedefaultmidpunct}
{\mcitedefaultendpunct}{\mcitedefaultseppunct}\relax
\EndOfBibitem
\bibitem[Doye and Calvo(2002)]{doye2002hsaLJ}
J.~P.~K. Doye and F.~Calvo, \emph{J. Chem. Phys.}, 2002, \textbf{116},
  8307--8317\relax
\mciteBstWouldAddEndPuncttrue
\mciteSetBstMidEndSepPunct{\mcitedefaultmidpunct}
{\mcitedefaultendpunct}{\mcitedefaultseppunct}\relax
\EndOfBibitem
\bibitem[Mandelshtam and Frantsuzov(2006)]{mandelshtam2006hsaLJ}
V.~A. Mandelshtam and P.~A. Frantsuzov, \emph{J. Chem. Phys.}, 2006,
  \textbf{124}, 204511\relax
\mciteBstWouldAddEndPuncttrue
\mciteSetBstMidEndSepPunct{\mcitedefaultmidpunct}
{\mcitedefaultendpunct}{\mcitedefaultseppunct}\relax
\EndOfBibitem
\bibitem[Sharapov and Mandelshtam(2007)]{sharapov2007hsaLJ}
V.~A. Sharapov and V.~A. Mandelshtam, \emph{J. Phys. Chem. A}, 2007,
  \textbf{111}, 10284--10291\relax
\mciteBstWouldAddEndPuncttrue
\mciteSetBstMidEndSepPunct{\mcitedefaultmidpunct}
{\mcitedefaultendpunct}{\mcitedefaultseppunct}\relax
\EndOfBibitem
\bibitem[Panizon and Ferrando(2015)]{panizon2015hsaPdPt}
E.~Panizon and R.~Ferrando, \emph{Phys. Rev. B}, 2015, \textbf{92},
  205417\relax
\mciteBstWouldAddEndPuncttrue
\mciteSetBstMidEndSepPunct{\mcitedefaultmidpunct}
{\mcitedefaultendpunct}{\mcitedefaultseppunct}\relax
\EndOfBibitem
\bibitem[Bonventre \emph{et~al.}(2018)Bonventre, Panizon, and
  Ferrando]{bonventre2018hsaAgCuANi}
D.~Bonventre, E.~Panizon and R.~Ferrando, \emph{Part. Part. Syst. Charact.},
  2018, \textbf{35}, 1700425\relax
\mciteBstWouldAddEndPuncttrue
\mciteSetBstMidEndSepPunct{\mcitedefaultmidpunct}
{\mcitedefaultendpunct}{\mcitedefaultseppunct}\relax
\EndOfBibitem
\bibitem[Rossi \emph{et~al.}(2004)Rossi, Rapallo, Mottet, Fortunelli, Baletto,
  and Ferrando]{rossi2004pIh}
G.~Rossi, A.~Rapallo, C.~Mottet, A.~Fortunelli, F.~Baletto and R.~Ferrando,
  \emph{Phys. Rev. Lett.}, 2004, \textbf{93}, 105503\relax
\mciteBstWouldAddEndPuncttrue
\mciteSetBstMidEndSepPunct{\mcitedefaultmidpunct}
{\mcitedefaultendpunct}{\mcitedefaultseppunct}\relax
\EndOfBibitem
\bibitem[Bracco \emph{et~al.}(1996)Bracco, Bruschi, Pedemonte, and
  Tatarek]{bracco1996Ag110AnharmonicRT}
G.~Bracco, L.~Bruschi, L.~Pedemonte and R.~Tatarek, \emph{Surf. Sci.}, 1996,
  \textbf{352--354}, 964--967\relax
\mciteBstWouldAddEndPuncttrue
\mciteSetBstMidEndSepPunct{\mcitedefaultmidpunct}
{\mcitedefaultendpunct}{\mcitedefaultseppunct}\relax
\EndOfBibitem
\bibitem[Bonella \emph{et~al.}(2012)Bonella, Meloni, and
  Ciccotti]{Bonella2012epjb}
S.~Bonella, S.~Meloni and G.~Ciccotti, \emph{The European Physical Journal B},
  2012, \textbf{85}, 97\relax
\mciteBstWouldAddEndPuncttrue
\mciteSetBstMidEndSepPunct{\mcitedefaultmidpunct}
{\mcitedefaultendpunct}{\mcitedefaultseppunct}\relax
\EndOfBibitem
\bibitem[Earl and Deem(2005)]{pearl2005PT}
D.~J. Earl and M.~W. Deem, \emph{Phys. Chem. Chem. Phys.}, 2005, \textbf{7},
  3910--3916\relax
\mciteBstWouldAddEndPuncttrue
\mciteSetBstMidEndSepPunct{\mcitedefaultmidpunct}
{\mcitedefaultendpunct}{\mcitedefaultseppunct}\relax
\EndOfBibitem
\bibitem[Neirotti \emph{et~al.}(2000)Neirotti, Calvo, Freeman, and
  Doll]{neirotti2000PTMCnLJ}
J.~P. Neirotti, F.~Calvo, D.~L. Freeman and J.~D. Doll, \emph{J. Chem. Phys.},
  2000, \textbf{112}, 10340--10349\relax
\mciteBstWouldAddEndPuncttrue
\mciteSetBstMidEndSepPunct{\mcitedefaultmidpunct}
{\mcitedefaultendpunct}{\mcitedefaultseppunct}\relax
\EndOfBibitem
\bibitem[Calvo \emph{et~al.}(2000)Calvo, Neirotti, Freeman, and
  Doll]{calvo2000PTMCnLJ}
F.~Calvo, J.~P. Neirotti, D.~L. Freeman and J.~D. Doll, \emph{J. Chem. Phys.},
  2000, \textbf{112}, 10350--10357\relax
\mciteBstWouldAddEndPuncttrue
\mciteSetBstMidEndSepPunct{\mcitedefaultmidpunct}
{\mcitedefaultendpunct}{\mcitedefaultseppunct}\relax
\EndOfBibitem
\bibitem[Noya and Doye(2006)]{noya2006PTMCnLJ}
E.~G. Noya and J.~P.~K. Doye, \emph{J. Chem. Phys.}, 2006, \textbf{124},
  104503\relax
\mciteBstWouldAddEndPuncttrue
\mciteSetBstMidEndSepPunct{\mcitedefaultmidpunct}
{\mcitedefaultendpunct}{\mcitedefaultseppunct}\relax
\EndOfBibitem
\bibitem[Cezar \emph{et~al.}(2017)Cezar, Rondina, and Silva]{cezar2017PTMCnLJ}
H.~M. Cezar, G.~G. Rondina and J.~L. F.~D. Silva, \emph{J. Chem. Phys.}, 2017,
  \textbf{146}, 064114\relax
\mciteBstWouldAddEndPuncttrue
\mciteSetBstMidEndSepPunct{\mcitedefaultmidpunct}
{\mcitedefaultendpunct}{\mcitedefaultseppunct}\relax
\EndOfBibitem
\bibitem[Senn \emph{et~al.}(2014)Senn, Wiebke, Schumann, and
  Gohr]{senn2014PTMCnAr}
F.~Senn, J.~Wiebke, O.~Schumann and S.~Gohr, \emph{J. Chem. Phys.}, 2014,
  \textbf{140}, 044325\relax
\mciteBstWouldAddEndPuncttrue
\mciteSetBstMidEndSepPunct{\mcitedefaultmidpunct}
{\mcitedefaultendpunct}{\mcitedefaultseppunct}\relax
\EndOfBibitem
\bibitem[Calvo(2008)]{calvo2008PTMCnPdPt}
F.~Calvo, \emph{Faraday Discuss.}, 2008, \textbf{138}, 75--88\relax
\mciteBstWouldAddEndPuncttrue
\mciteSetBstMidEndSepPunct{\mcitedefaultmidpunct}
{\mcitedefaultendpunct}{\mcitedefaultseppunct}\relax
\EndOfBibitem
\bibitem[Calvo \emph{et~al.}(2008)Calvo, Cottancin, and
  Broyer]{calvo2008PTMCnAgM}
F.~Calvo, E.~Cottancin and M.~Broyer, \emph{Phys. Rev. B}, 2008, \textbf{77},
  121406(R)\relax
\mciteBstWouldAddEndPuncttrue
\mciteSetBstMidEndSepPunct{\mcitedefaultmidpunct}
{\mcitedefaultendpunct}{\mcitedefaultseppunct}\relax
\EndOfBibitem
\bibitem[Calvo and Mottet(2011)]{calvo2011PTMCnCoPt}
F.~Calvo and C.~Mottet, \emph{Phys. Rev. B}, 2011, \textbf{84}, 035409\relax
\mciteBstWouldAddEndPuncttrue
\mciteSetBstMidEndSepPunct{\mcitedefaultmidpunct}
{\mcitedefaultendpunct}{\mcitedefaultseppunct}\relax
\EndOfBibitem
\bibitem[Cezar \emph{et~al.}(2019)Cezar, Rondina, and
  Silva]{cezar2019PTMCnPtCoNi}
H.~M. Cezar, G.~G. Rondina and J.~L. F.~D. Silva, \emph{J. Chem. Phys.}, 2019,
  \textbf{151}, 204301\relax
\mciteBstWouldAddEndPuncttrue
\mciteSetBstMidEndSepPunct{\mcitedefaultmidpunct}
{\mcitedefaultendpunct}{\mcitedefaultseppunct}\relax
\EndOfBibitem
\bibitem[Shu \emph{et~al.}(2012)Shu, Yang, Zhai, Sun, Xiang, and
  Gong]{calvo2012PTMDnFeMelting}
Q.~Shu, Y.~Yang, Y.~Zhai, D.~Y. Sun, H.~J. Xiang and X.~G. Gong,
  \emph{Nanoscale}, 2012, \textbf{4}, 6307--6311\relax
\mciteBstWouldAddEndPuncttrue
\mciteSetBstMidEndSepPunct{\mcitedefaultmidpunct}
{\mcitedefaultendpunct}{\mcitedefaultseppunct}\relax
\EndOfBibitem
\bibitem[Tarrat \emph{et~al.}(2018)Tarrat, Rapacioli, and
  Spiegelman]{tarrat2018PTMDnAu}
N.~Tarrat, M.~Rapacioli and F.~Spiegelman, \emph{J. Chem. Phys.}, 2018,
  \textbf{148}, 204308\relax
\mciteBstWouldAddEndPuncttrue
\mciteSetBstMidEndSepPunct{\mcitedefaultmidpunct}
{\mcitedefaultendpunct}{\mcitedefaultseppunct}\relax
\EndOfBibitem
\bibitem[Wales and Doye(1997)]{Wales1997jpca}
D.~J. Wales and J.~P.~K. Doye, \emph{J. Phys. Chem. A}, 1997, \textbf{101},
  5111--5116\relax
\mciteBstWouldAddEndPuncttrue
\mciteSetBstMidEndSepPunct{\mcitedefaultmidpunct}
{\mcitedefaultendpunct}{\mcitedefaultseppunct}\relax
\EndOfBibitem
\bibitem[Sharapov \emph{et~al.}(2007)Sharapov, Meluzzi, and
  Mandelshtam]{sharapov2007combinePTMDnHSA}
V.~A. Sharapov, D.~Meluzzi and V.~A. Mandelshtam, \emph{Phys. Rev. Lett.},
  2007, \textbf{98}, 105701\relax
\mciteBstWouldAddEndPuncttrue
\mciteSetBstMidEndSepPunct{\mcitedefaultmidpunct}
{\mcitedefaultendpunct}{\mcitedefaultseppunct}\relax
\EndOfBibitem
\bibitem[Bogdan and Wales(2006)]{bogdan2006hsaBasinSampling}
T.~V. Bogdan and D.~J. Wales, \emph{J. Chem. Phys.}, 2006, \textbf{124},
  044102\relax
\mciteBstWouldAddEndPuncttrue
\mciteSetBstMidEndSepPunct{\mcitedefaultmidpunct}
{\mcitedefaultendpunct}{\mcitedefaultseppunct}\relax
\EndOfBibitem
\bibitem[Ballard and Wales(2014)]{ballard2014superposOptTemp}
A.~J. Ballard and D.~J. Wales, \emph{J. Chem. Theory Comput.}, 2014,
  \textbf{10}, 5599--5605\relax
\mciteBstWouldAddEndPuncttrue
\mciteSetBstMidEndSepPunct{\mcitedefaultmidpunct}
{\mcitedefaultendpunct}{\mcitedefaultseppunct}\relax
\EndOfBibitem
\bibitem[Cyrot-Lackmann and Ducastelle(1971)]{tbsma1971}
F.~Cyrot-Lackmann and F.~Ducastelle, \emph{Phys. Rev. B: Condens. Matter Mater.
  Phys.}, 1971, \textbf{4}, 2406--2412\relax
\mciteBstWouldAddEndPuncttrue
\mciteSetBstMidEndSepPunct{\mcitedefaultmidpunct}
{\mcitedefaultendpunct}{\mcitedefaultseppunct}\relax
\EndOfBibitem
\bibitem[Baletto \emph{et~al.}(2002)Baletto, Ferrando, Fortunelli, Montalent,
  and Mottet]{baletto2002potParams}
F.~Baletto, R.~Ferrando, A.~Fortunelli, F.~Montalent and C.~Mottet, \emph{J.
  Chem. Phys.}, 2002, \textbf{116}, 3856--3863\relax
\mciteBstWouldAddEndPuncttrue
\mciteSetBstMidEndSepPunct{\mcitedefaultmidpunct}
{\mcitedefaultendpunct}{\mcitedefaultseppunct}\relax
\EndOfBibitem
\bibitem[Han \emph{et~al.}(2014)Han, Ferrando, and Li]{han2014imagingOnMgO}
Y.~Han, R.~Ferrando and Z.~Y. Li, \emph{J. Phys. Chem. Lett.}, 2014,
  \textbf{5}, 131--137\relax
\mciteBstWouldAddEndPuncttrue
\mciteSetBstMidEndSepPunct{\mcitedefaultmidpunct}
{\mcitedefaultendpunct}{\mcitedefaultseppunct}\relax
\EndOfBibitem
\bibitem[Apra \emph{et~al.}(2004)Apra, Baletto, Ferrando, and
  Fortunelli]{apra2004AuRosette}
E.~Apra, F.~Baletto, R.~Ferrando and A.~Fortunelli, \emph{Phys. Rev. Lett.},
  2004, \textbf{93}, 065502\relax
\mciteBstWouldAddEndPuncttrue
\mciteSetBstMidEndSepPunct{\mcitedefaultmidpunct}
{\mcitedefaultendpunct}{\mcitedefaultseppunct}\relax
\EndOfBibitem
\bibitem[Guillope and Legrand(1989)]{guillope1989surfRecons}
M.~Guillope and B.~Legrand, \emph{Surf. Sci.}, 1989, \textbf{215},
  577--595\relax
\mciteBstWouldAddEndPuncttrue
\mciteSetBstMidEndSepPunct{\mcitedefaultmidpunct}
{\mcitedefaultendpunct}{\mcitedefaultseppunct}\relax
\EndOfBibitem
\bibitem[Ferrando and Treglia(1995)]{ferrando1995surfDiffusion}
R.~Ferrando and G.~Treglia, \emph{Surf. Sci.}, 1995, \textbf{331--333},
  920--924\relax
\mciteBstWouldAddEndPuncttrue
\mciteSetBstMidEndSepPunct{\mcitedefaultmidpunct}
{\mcitedefaultendpunct}{\mcitedefaultseppunct}\relax
\EndOfBibitem
\bibitem[Boisvert \emph{et~al.}(1995)Boisvert, Lewis, Puska, and
  Nieminen]{boisvert1995surfDiffusionDFT}
G.~Boisvert, L.~J. Lewis, M.~J. Puska and R.~M. Nieminen, \emph{Phys. Rev. B},
  1995, \textbf{52}, 9078--9085\relax
\mciteBstWouldAddEndPuncttrue
\mciteSetBstMidEndSepPunct{\mcitedefaultmidpunct}
{\mcitedefaultendpunct}{\mcitedefaultseppunct}\relax
\EndOfBibitem
\bibitem[Rossi and Ferrando(2009)]{rossi2009BHMC}
G.~Rossi and R.~Ferrando, \emph{J. Phys.: Condens. Matter}, 2009, \textbf{21},
  084208\relax
\mciteBstWouldAddEndPuncttrue
\mciteSetBstMidEndSepPunct{\mcitedefaultmidpunct}
{\mcitedefaultendpunct}{\mcitedefaultseppunct}\relax
\EndOfBibitem
\bibitem[Nelli and Ferrando(2019)]{Nelli2019nanoscale}
D.~Nelli and R.~Ferrando, \emph{Nanoscale}, 2019, \textbf{11},
  13040--13050\relax
\mciteBstWouldAddEndPuncttrue
\mciteSetBstMidEndSepPunct{\mcitedefaultmidpunct}
{\mcitedefaultendpunct}{\mcitedefaultseppunct}\relax
\EndOfBibitem
\bibitem[Lingenheil \emph{et~al.}(2009)Lingenheil, Denschlag, Mathias, and
  Tavan]{lingenheil2009DEOSEO}
M.~Lingenheil, R.~Denschlag, G.~Mathias and P.~Tavan, \emph{Chem. Phys. Lett},
  2009, \textbf{478}, 80--84\relax
\mciteBstWouldAddEndPuncttrue
\mciteSetBstMidEndSepPunct{\mcitedefaultmidpunct}
{\mcitedefaultendpunct}{\mcitedefaultseppunct}\relax
\EndOfBibitem
\bibitem[Plimpton(1995)]{lammps}
S.~Plimpton, \emph{J. Comp. Phys.}, 1995, \textbf{117}, 1--19\relax
\mciteBstWouldAddEndPuncttrue
\mciteSetBstMidEndSepPunct{\mcitedefaultmidpunct}
{\mcitedefaultendpunct}{\mcitedefaultseppunct}\relax
\EndOfBibitem
\bibitem[Faken and Jonssonn(1994)]{cna1994}
D.~Faken and H.~Jonssonn, \emph{Comput. Mater. Sci.}, 1994, \textbf{2},
  279--286\relax
\mciteBstWouldAddEndPuncttrue
\mciteSetBstMidEndSepPunct{\mcitedefaultmidpunct}
{\mcitedefaultendpunct}{\mcitedefaultseppunct}\relax
\EndOfBibitem
\bibitem[Bedoya-Martinez \emph{et~al.}(2006)Bedoya-Martinez, Kaczmarski, and
  Hernandez]{guptaPotUnderEstMP2006}
O.~N. Bedoya-Martinez, M.~Kaczmarski and E.~R. Hernandez, \emph{J. Condens.
  Matter Phys.}, 2006, \textbf{18}, 8049--8062\relax
\mciteBstWouldAddEndPuncttrue
\mciteSetBstMidEndSepPunct{\mcitedefaultmidpunct}
{\mcitedefaultendpunct}{\mcitedefaultseppunct}\relax
\EndOfBibitem
\bibitem[Hartigan and Wong(1979)]{kmeans1979}
J.~A. Hartigan and M.~A. Wong, \emph{J R Stat. Soc. Ser. C Appl.}, 1979,
  \textbf{28}, 100--108\relax
\mciteBstWouldAddEndPuncttrue
\mciteSetBstMidEndSepPunct{\mcitedefaultmidpunct}
{\mcitedefaultendpunct}{\mcitedefaultseppunct}\relax
\EndOfBibitem
\bibitem[Lloyd(1982)]{kmeans1982}
S.~P. Lloyd, \emph{IEEE Trans. Inf. Theory}, 1982, \textbf{28}, 129--137\relax
\mciteBstWouldAddEndPuncttrue
\mciteSetBstMidEndSepPunct{\mcitedefaultmidpunct}
{\mcitedefaultendpunct}{\mcitedefaultseppunct}\relax
\EndOfBibitem
\bibitem[Rossi and Ferrando(2007)]{polyDh2007}
G.~Rossi and R.~Ferrando, \emph{Nanotechnology}, 2007, \textbf{18},
  225706\relax
\mciteBstWouldAddEndPuncttrue
\mciteSetBstMidEndSepPunct{\mcitedefaultmidpunct}
{\mcitedefaultendpunct}{\mcitedefaultseppunct}\relax
\EndOfBibitem
\bibitem[Kostko \emph{et~al.}(2007)Kostko, Huber, Moseler, and von
  Issendorff]{kostko2007chiral71}
O.~Kostko, B.~Huber, M.~Moseler and B.~von Issendorff, \emph{Phys. Rev. Lett.},
  2007, \textbf{98}, 043401\relax
\mciteBstWouldAddEndPuncttrue
\mciteSetBstMidEndSepPunct{\mcitedefaultmidpunct}
{\mcitedefaultendpunct}{\mcitedefaultseppunct}\relax
\EndOfBibitem
\bibitem[Rapallo \emph{et~al.}(2012)Rapallo, Olmos-Asar, Oviedo, Ludena,
  Ferrando, and Mariscal]{rapallo2012chiral71AuCo}
A.~Rapallo, J.~A. Olmos-Asar, O.~A. Oviedo, M.~Ludena, R.~Ferrando and M.~M.
  Mariscal, \emph{J. Phys. Chem. C}, 2012, \textbf{116}, 17210--17218\relax
\mciteBstWouldAddEndPuncttrue
\mciteSetBstMidEndSepPunct{\mcitedefaultmidpunct}
{\mcitedefaultendpunct}{\mcitedefaultseppunct}\relax
\EndOfBibitem
\end{mcitethebibliography}
\bibliographystyle{rsc} 

\end{document}